\documentclass[%
 reprint,
superscriptaddress,
amsmath,amssymb,twocolumn,
aps,longbibliography
]{revtex4-1}
\usepackage{graphicx}
\usepackage{graphics}
\usepackage[pdftex]{color}
\usepackage{xcolor}
\usepackage{dcolumn}
\usepackage{bm}
\usepackage{siunitx}
\usepackage{xr}
\usepackage{amsmath}
\usepackage{float}

\begin{document}
\title{Probing the screening of the Casimir interaction with optical tweezers}
\author{L. B. Pires} 
\affiliation{LPO-COPEA, Instituto de Ci\^encias Biom\'edicas, Universidade Federal do Rio de Janeiro, Rio de Janeiro, RJ, 21941-902, Brazil}
\affiliation{CENABIO - Centro Nacional de Biologia Estrutural e Bioimagem, Universidade Federal do Rio de Janeiro,
Rio de Janeiro, Rio de Janeiro, 21941-902, Brazil}
\affiliation{Instituto de F\'{\i}sica, Universidade Federal do Rio de Janeiro, Caixa Postal 68528,   Rio de Janeiro,  RJ, 21941-972, Brazil}
\author{D. S. Ether} 
\affiliation{LPO-COPEA, Instituto de Ci\^encias Biom\'edicas, Universidade Federal do Rio de Janeiro, Rio de Janeiro, RJ, 21941-902, Brazil}
\affiliation{CENABIO - Centro Nacional de Biologia Estrutural e Bioimagem, Universidade Federal do Rio de Janeiro,
Rio de Janeiro, Rio de Janeiro, 21941-902, Brazil}
\affiliation{Instituto de F\'{\i}sica, Universidade Federal do Rio de Janeiro, Caixa Postal 68528,   Rio de Janeiro,  RJ, 21941-972, Brazil}
\author{B. Spreng} 
\affiliation{Universit\"at Augsburg, Institut f\"ur Physik, 86135 Augsburg, Germany}
\affiliation{Department of Electrical and Computer Engineering, University of California, Davis, California 95616, USA}
\author{G. R. S. Ara\'ujo} 
\affiliation{Instituto de Biof\'isica Carlos Chagas Filho, Rio de Janeiro, Rio de Janeiro, 21941-901, Brazil}
\author{R. S. Decca} 
\affiliation{Department of Physics, Indiana University-Purdue University Indianapolis, Indianapolis, Indiana 46202, USA}
\author{R. S. Dutra} 
\affiliation{LISComp-IFRJ, Instituto Federal de Educação, Ciência e Tecnologia, Rua Sebastão de Lacerda, Paracambi, RJ 26600-000, Brasil}
\author{M. Borges} 
\affiliation{LPO-COPEA, Instituto de Ci\^encias Biom\'edicas, Universidade Federal do Rio de Janeiro, Rio de Janeiro, RJ, 21941-902, Brazil}
\affiliation{CENABIO - Centro Nacional de Biologia Estrutural e Bioimagem, Universidade Federal do Rio de Janeiro,
Rio de Janeiro, Rio de Janeiro, 21941-902, Brazil}
\affiliation{Instituto de F\'{\i}sica, Universidade Federal do Rio de Janeiro, Caixa Postal 68528,   Rio de Janeiro,  RJ, 21941-972, Brazil}
\author{F. S. S. Rosa} 
\affiliation{Instituto de F\'{\i}sica, Universidade Federal do Rio de Janeiro, Caixa Postal 68528,   Rio de Janeiro,  RJ, 21941-972, Brazil}
\author{G.-L. Ingold} 
\affiliation{Universit\"at Augsburg, Institut f\"ur Physik, 86135 Augsburg, Germany}
\author{M. J. B. Moura} 
\affiliation{Departamento de Engenharia Mec\^anica, Pontif\'icia Universidade Cat\'olica do Rio de Janeiro, RJ, 22451-900, Brazil}
\author{S. Frases} 
\affiliation{Instituto de Biof\'isica Carlos Chagas Filho, Rio de Janeiro, Rio de Janeiro, 21941-901, Brazil}
\author{B. Pontes} 
\affiliation{LPO-COPEA, Instituto de Ci\^encias Biom\'edicas, Universidade Federal do Rio de Janeiro, Rio de Janeiro, RJ, 21941-902, Brazil}
\affiliation{CENABIO - Centro Nacional de Biologia Estrutural e Bioimagem, Universidade Federal do Rio de Janeiro,
Rio de Janeiro, Rio de Janeiro, 21941-902, Brazil}
\affiliation{Instituto de Ci\^encias Biom\'edicas, Universidade Federal do Rio de Janeiro, Rio de Janeiro, RJ 21941-902, Brazil}
\author{H. M. Nussenzveig}
\affiliation{LPO-COPEA, Instituto de Ci\^encias Biom\'edicas, Universidade Federal do Rio de Janeiro, Rio de Janeiro, RJ, 21941-902, Brazil}
\affiliation{CENABIO - Centro Nacional de Biologia Estrutural e Bioimagem, Universidade Federal do Rio de Janeiro,
Rio de Janeiro, Rio de Janeiro, 21941-902, Brazil}
\affiliation{Instituto de F\'{\i}sica, Universidade Federal do Rio de Janeiro, Caixa Postal 68528,   Rio de Janeiro,  RJ, 21941-972, Brazil}
\author{S. Reynaud}
\affiliation{Laboratoire Kastler Brossel, Sorbonne Universit\'e, CNRS, ENS-PSL Universit\'e, Coll\`ege de France,
Campus Pierre et Marie Curie, 75252 Paris, France}
\author{N. B. Viana}
\affiliation{LPO-COPEA, Instituto de Ci\^encias Biom\'edicas, Universidade Federal do Rio de Janeiro, Rio de Janeiro, RJ, 21941-902, Brazil}
\affiliation{CENABIO - Centro Nacional de Biologia Estrutural e Bioimagem, Universidade Federal do Rio de Janeiro,
Rio de Janeiro, Rio de Janeiro, 21941-902, Brazil}
\affiliation{Instituto de F\'{\i}sica, Universidade Federal do Rio de Janeiro, Caixa Postal 68528,   Rio de Janeiro,  RJ, 21941-972, Brazil}
\author{P. A. Maia Neto}
\affiliation{LPO-COPEA, Instituto de Ci\^encias Biom\'edicas, Universidade Federal do Rio de Janeiro, Rio de Janeiro, RJ, 21941-902, Brazil}
\affiliation{CENABIO - Centro Nacional de Biologia Estrutural e Bioimagem, Universidade Federal do Rio de Janeiro,
Rio de Janeiro, Rio de Janeiro, 21941-902, Brazil}
\affiliation{Instituto de F\'{\i}sica, Universidade Federal do Rio de Janeiro, Caixa Postal 68528,   Rio de Janeiro,  RJ, 21941-972, Brazil}
\date{\today}
\begin{abstract}

 We measure the colloidal interaction between two silica microspheres in aqueous solution in the distance range from  $0.2\,\mu\si{m}$ to $0.5\,\mu\si{m}$
 with the help of optical tweezers. When employing a sample with a low salt concentration, the resulting interaction is dominated by the repulsive double-layer interaction
which is fully characterized. 
The double-layer interaction is suppressed when adding $0.22 ~\si{M}$ of salt to our sample,  thus leading to a purely attractive Casimir signal.
When analyzing the experimental data for the potential energy and force, 
we find good agreement with 
theoretical results based on the scattering approach. At the distance range probed experimentally, the
 interaction arises mainly from the unscreened transverse magnetic contribution in the zero-frequency limit, with nonzero Matsubara frequencies providing a negligible contribution.
 In contrast, such unscreened contribution is not included by the standard theoretical model of the Casimir interaction in electrolyte solutions, in which
 the zero-frequency term is treated separately as an electrostatic fluctuational effect. 
 As a consequence, the resulting attraction is too weak in this standard model, by approximately one order of magnitude, to explain the experimental data. 
 Overall, our experimental results shed light on the nature of the thermal zero-frequency contribution and indicate that the Casimir attraction across polar liquids has a longer range than previously predicted. 
 
\end{abstract}
\maketitle
\section{Introduction}

The van der Waals (vdW) interaction plays a key role in several 
systems at the intersection between cell and molecular biology, chemistry and physics,
such as biological membranes and
colloids, among others \cite{israelachvili2011intermolecular,butt10}.
The general theoretical framework for the vdW force was laid down by Lifshitz~\cite{Lifshitz1956}, whose work was later extended to allow for 
an intervening material medium between the interacting surfaces~\cite{dzyaloshinskii1961general}. 
The vdW interaction across an electrolyte solution is a key ingredient of the Derjaguin, Landau, Verwey, and Overbeek (DLVO) theory of colloidal interactions \cite{russel1991colloidal}.
Electrodynamic retardation leads to a reduction of the contribution 
from nonzero Matsubara frequencies as the distance increases beyond the nanometric scale.
The vdW interaction is often referred to as Casimir interaction~\cite{Casimir1948} in this range, and several
 experiments with metallic surfaces either in vacuum~\cite{Lamoreaux1997,Mohideen1998,Chan2001,Decca2007,Sushkov2011,Chang2012,Bimonte2016,Liu2019,Bimonte2021} or in 
 air~\cite{deMan2009,Garret2018} have been reported
(see~\cite{Klimchitskaya2020,Gong2021} for recent reviews). 

In many situations of interest, 
the asymptotic long-distance Casimir interaction arises from the Matsubara zero frequency, which 
provides a purely thermal contribution proportional to the temperature $T.$  
However, when considering surfaces separated by an electrolyte solution, screening by movable ions
(partly) suppresses the zero-frequency contribution over the characteristic Debye screening length~\cite{Woods2016}, while leaving the contribution of nonzero frequencies unchanged~\cite{neto2019scattering}. 

The screening of the vdW interaction has been demonstrated for complex systems such as lipid bi-layers~\cite{petrache2006salt},
for which a direct comparison with ab-initio theoretical models is not available.
We have employed optical tweezers \cite{Ashkin86,ashkin06,gennerich17} to probe the interaction between silica microspheres
across an aqueous solution
 at surface-to-surface distances $L\gtrsim 0.2\,\mu{\rm m}$ such that the
thermal zero-frequency contribution dominates the Casimir signal. 
Under such conditions, the Casimir interaction is extremely sensitive to the strength of ionic screening.
In addition, the simplicity of our setup allows for a direct comparison with theoretical descriptions of screening. 

At such distances, the magnitude of the Casimir force between dielectric surfaces
 is in the femtonewton (fN) range, thus requiring the use of very soft tweezers, with a stiffness of the order
of fN/nm and long measurement times.  Such force sensitivity was achieved by carefully mitigating the laser and 
microscope stage drifts, non-thermal noises and 
spurious measurement signals arising from perturbations of the optical potential by the presence of the 
additional interacting microsphere. 

A much stronger interaction is obtained by replacing dielectric surfaces by metallic ones.
Experiments with different intervening liquids
were implemented in the distance range close to $\sim 0.1\,\mu{\rm m}$~\cite{munday07,munday08,munday09,Zwol2010,LeCunuder2018}.
However, in this case the zero-frequency contribution becomes dominant only at distances 
  above the micrometer range. Thus, 
  it would be extremely difficult to probe the screening effect with  metallic surfaces. 
In contrast, screening is relevant already in the nanometer range when considering several examples of dielectric surfaces interacting across 
 an aqueous medium. Indeed, in this case the zero-frequency contribution typically stands out as the dominant term already at relatively short distances, of the order 
 of  $\sim 0.1\,\mu{\rm m},$ due to the near index-matching 
 at nonzero Matsubara frequencies~\cite{Parsegian1971,russel1991colloidal}. 
For instance, the vdW interaction between 
 lipid membranes in the nanometer range is strongly modified by changing the salt concentration and the resulting screening length~\cite{petrache2006salt}. 
Thus, optical tweezers are ideally suited for probing the Casimir screening 
as it allows for weak trapping of dielectric particles in aqueous solution as long as 
the requirement of near index matching is satisfied \cite{ashkin}, 
which is precisely the condition for making the zero-frequency Casimir contribution dominant in the sub-micron range. 

Previously, optical tweezers were employed as force transducers 
to probe colloidal forces \cite{schaffer07,gutsche2007forces,elmahdy2010forces,dominguez2008optical,
hansen2005novel,ether15,kundu2019measurement} and the
 non-additivity of the critical Casimir interaction~\cite{paladugu2016nonadditivity}. Alternatively,
 one can use blinking optical tweezers to control the initial distance between interacting dielectric microspheres \cite{crocker1994microscopic,grier1997optical,sainis2007statistics,sainis08,polin2007colloidal}. Recently, femtonewton force sensitivity was achieved using an optically trapped metallic nanosphere as the probe~\cite{schnoering19}.
When combined with total internal reflection microscopy (TIRM)~\cite{prieve1990}, optical tweezers allow for subfemtonewton measurements~\cite{liu16} 
when reflection of the probe laser at the planar interface between the aqueous medium and the substrate is reduced~\cite{liu14}.
Potential energy measurements of the critical~\cite{Hertlein2008} and 
electrodynamic \cite{bevan1999,nayeri2013measurements,Cao2019} Casimir interactions were implemented using TIRM. 
Atomic force microscopy (AFM) is usually the method of choice when probing at shorter distances (see \cite{smith2020forces}
for a recent review). Typically, AFM cantilevers with stiffness $\sim \si{pN/nm}$ allow to measure piconewton forces at distances in the nanometer range~\cite{wodka14}.

With the notable exceptions of Refs.~\cite{hansen2005novel,kundu2019measurement}, measurements
with dielectric surfaces separated by distances $\gtrsim  0.1\,\mu{\rm m}$ are typically made with low or intermediate salt concentrations, with
the electrostatic double layer repulsive force providing a significant fraction of the total interaction. 
We have probed samples with low and high salt concentrations. For the latter, 
the electrostatic double-layer force is completely suppressed, thus allowing for a direct blind comparison between the experimental data and theoretical models for 
the Casimir effect. 
Measuring at such conditions at distances $\gtrsim 0.2\,\mu{\rm m}$ is challenging not only because the 
 signal is weak, but also because the steep attractive Casimir potential leads to frequent jump-into-contact events. 

We compare our experimental results with two different models of screening of the Casimir interaction. In the first one, the zero-frequency contribution is considered separately within the realm of fluctuational electrostatics taking the ions into account with the help of  the linear Poisson-Boltzmann equation~\cite{Mitchell1974,Mahanty1976,parsegian2005van}. Similarly to the double-layer interaction between charged surfaces, the resulting interaction is suppressed over the screening length.

The second model is based on 
 a recent extension~\cite{neto2019scattering} of the scattering approach for parallel planar surfaces~\cite{Jaekel1991,Genet2003} to include longitudinal channels. 
In contrast to the former standard model~\cite{Mitchell1974,Mahanty1976,parsegian2005van}, a single formalism is 
applied to all Matsubara frequencies, and the zero-frequency contribution is obtained as a limit of the result for an arbitrary frequency. 
Movable ions in solution give rise to a nonlocal response (spatial dispersion), which in
turn allows for the existence of longitudinal modes in addition to the usual transverse ones~\cite{davies1972van}. 

The effect of the nonlocal ionic response is shown to be negligible
at nonzero Matsubara frequencies, because they are much larger than the plasma frequency
associated to ions in solution~\cite{neto2019scattering}. 
As for the zero-frequency contribution, 
the known screened term of Refs.~\cite{Mitchell1974,Mahanty1976,parsegian2005van} is re-derived as the contribution of longitudinal channels.
However,
an additional unscreened contribution is also obtained within the scattering model,
 resulting from transverse magnetic (TM) modes in the
zero-frequency limit. As a result of the TM contribution, the 
Casimir interaction across a salt solution is predicted to be of a much longer range than previously thought, 
corresponding to a universal long-distance asymptotic 
Hamaker constant $\sim 0.9\,k_B T$ in the case of parallel planar surfaces ($k_B=$ Boltzmann constant). 
Our experimental data at a high salt concentration allow us to check for the existence of this 
additional TM contribution, which 
is one order of magnitude larger than the contribution of nonzero frequencies 
in the probed distance range $L>0.2\,\mu{\rm m}.$  

The paper is organized as follows. Sec.~II presents the basic ingredients 
for the theoretical description of the colloidal interaction between dielectric microspheres. 
Sec.~III starts with a description of the experiment and then 
presents a detailed comparison between results and theoretical models. Sec.~IV contains the conclusions and the final remarks. More technical aspects of the experimental methods and additional data are presented 
in appendices A to E.   

\section{\label{sec:inte}Theory of Colloidal Interactions}

Here we consider the interaction between two silica microspheres. 
While the probe microsphere (radius $R_1$) is optically trapped, a larger one (radius $R_2$) is adhered to the coverslip, as illustrated by Fig.~\ref{fig:1}(a). 
They are separated by a surface-to-surface distance $L$ in an aqueous solution of permittivity $\epsilon$ at temperature $T$. For the distance range $L\gtrsim 0.2\,\mu{\rm m}$ probed in our experiment, non-DLVO short-range interactions such as 
solvation forces are negligible~\cite{donaldson15}, and the total interaction energy between the microspheres reads
\begin{equation}
U_{\rm int}(L) =  U_{\rm DL}(L)+ U_{\rm C}(L)\, ,
\end{equation}
where $ U_{\rm DL}$ and $U_{\rm C}$ are the double layer and Casimir interaction energies, respectively. 
The Debye screening length~\cite{israelachvili2011intermolecular,butt10}
\begin{equation}
\lambda_{\rm D}=\sqrt{\frac{\epsilon k_{B}T}{2(Ze)^2n_{\infty}}}
\label{eq.lambdaD}
\end{equation}
is the characteristic thickness of the ionic double layer around each particle. Here,  
 $k_{B}$ denotes the Boltzmann constant, $Z$ is the ion atomic number and $e$ is the elementary charge.
The Debye length and hence the screening intensity can be controlled by changing
the bulk salt concentration $n_{\infty}.$ 

In the regime of long distances, the double layer around each microsphere is 
approximately unperturbed by the other one. 
Within the  linear superposition approximation (LSA),
the ionic charge density is then taken as the sum of separate solutions of the linear Poisson-Boltzmann (Debye-H\"uckel) equation
for isolated microspheres~\cite{bell1970approximate}.
Indeed, the relative difference between the LSA result and the exact solution of the linear Poisson-Boltzmann equation for two spheres is negligibly small given our experimental conditions \cite{Ether2018}.
 As charge regulation is also
negligible at distances $L> \lambda_{\rm D}$ \cite{carnie93A,trefalt16},
we consider a constant charge model, with the same fixed surface charge density $\sigma$ on both silica microspheres. 
The resulting screened Coulomb interaction energy is then given by~\cite{carnie94}
\begin{equation}
U_{\rm DL}(L)=\frac{4\pi\sigma^{2}R_{1}R_{2}}{\epsilon\left( 1+\frac{R_{1}}{\lambda_{\rm D}}\right)\left(1+\frac{R_{2}}{\lambda_{\rm D}}\right)} \frac{R_{\rm eff}}{\left(1+\frac{L}{R_{1}+R_{2}}\right)} \textrm{e}^{-L/\lambda_{\rm D}}\, ,
\label{eq.dcLSA}
\end{equation}  
where $R_{\rm eff}\equiv R_{1}R_{2}/(R_{1}+R_{2})$ is the effective radius.

The Casimir interaction is analyzed within the scattering approach to non-trivial geometries \cite{lambrecht2006casimir,Rahi2009},
which allows the derivation of exact results for the spherical geometry in terms of the corresponding Mie scattering operators~\cite{Emig2007,MaiaNeto2008,Canaguier-Durand2010,Rodriguez-Lopez2011} developed in the plane-wave basis~\cite{spreng2020}. 
Our approach also accounts for
electrodynamic retardation and thermal effects. 
As the distance increases, deviations from the standard proximity force approximation (PFA), 
also known as Derjaguin approximation~\cite{Derjaguin1934},
become increasingly important~\cite{MaiaNeto2008,Emig2008,Teo2011,Bimonte2012,Bimonte2012EPL,Hartmann2017,Bimonte2018b,Henning2019}.

The Casimir free energy 
is then given as a sum over the Matsubara frequencies 
\begin{equation}
 \xi_n=2\pi\, n\, k_B
T/\hbar,\;n=0,1,2,...
\end{equation}
that reads
\begin{equation}
\label{scatteringformula}
U_{\rm C}(L) = k_B T \sum_{n=0}^\infty{}'
              \log\det\left[1-\mathcal{M}(\xi_n)\right].
\end{equation}
The prime stands for an addition factor $1/2$
when considering the zero-frequency 
($n=0$) contribution.
The round-trip operator 
\begin{equation}\label{eq:roundtrip_operator}
\mathcal{M}(\xi_n) =  \mathcal{T}_{12}\mathcal{R}_2 \mathcal{T}_{21}\mathcal{R}_1
\end{equation}
 contains
 the reflection (Mie) operators
  $\mathcal{R}_1$ and $\mathcal{R}_2$ 
of the probe and adhered microsphere, respectively. 
 They are calculated with respect to reference points located at their corresponding centers. 
The operator $\mathcal{T}_{21}$ carries out the
translation from the center of microsphere $1$ to the one of microsphere $2$ along the $x$ direction.
Likewise, $\mathcal{T}_{12}$ implements the opposite translation from microsphere $2$ to $1,$
also across the center-to-center distance $L+R_1+R_2,$ as illustrated by Fig.~\ref{fig:1}(a).

We follow \cite{spreng2020} and develop the scattering formula (\ref{scatteringformula}) in the plane-wave basis with the help of a discrete Fourier transform. 
We take the dielectric functions of silica and water from \cite{Zwol2010}. The contribution to the medium dielectric function 
arising from ions in solution is considered within the Drude model, and is non-negligible only for the zero-frequency contribution. 

As discussed in the previous section, the screening of 
the Casimir interaction by movable ions in solution is the central focus of the present paper. 
 For the simpler geometry of planar parallel surfaces, the scattering formula for the Casimir interaction across an electrolyte solution was recently developed in terms of 
 the nonlocal electrodynamic response of the intervening medium~\cite{neto2019scattering}. 
 Only the zero-frequency contribution is modified by ions in solution, as the corresponding plasma frequency is much smaller than $k_BT/\hbar.$
 Two separate contributions were found at the zero-frequency limit: 
 The first one, accounting for longitudinal modes, coincides with the result of previous derivations based on the linear Poisson-Boltzmann equation~\cite{Mitchell1974,Mahanty1976,parsegian2005van}.
 Alongside the screened longitudinal term, the scattering approach of Ref.~\cite{neto2019scattering}
  leads to an additional, unscreened contribution arising from TM modes in the zero-frequency limit. 

The main purpose of the present study is to investigate whether such additional contribution is found experimentally. 
 We compare our data with the full Mie calculation based on the scattering formula (\ref{scatteringformula}) for spherical particles
 instead of directly applying the results of \cite{neto2019scattering}
 or~\cite{Mitchell1974,Mahanty1976,parsegian2005van} for parallel planar surfaces.
 Since we probe distances $L\gg \lambda_{\rm D},$ the zero-frequency contribution in (\ref{scatteringformula}) is completely suppressed 
if we follow the general scheme of Refs.~\cite{Mitchell1974,Mahanty1976,parsegian2005van}. Indeed, 
in this case the zero-frequency contribution is considered separately as an electrostatic effect derived from the linear Poisson-Boltzmann equation, and then
all multipole contributions are screened over the Debye length $\lambda_{\rm D}$~\cite{Renan2021}. 
On the other hand, when taking the zero-frequency contribution as a limit of the general scattering formalism, an additional 
 unscreened TM contribution provides the long-distance
 asymptotic value of the Casimir attraction, as the contribution of nonzero Matsubara frequencies 
 becomes negligible at the distances
 $L\gtrsim 0.2\,\mu{\rm m}$
 probed in our experiment. 
 Such asymptotic result is of a universal nature as it does not depend on the details of the dielectric functions of the 
 materials involved in the experiment~\cite{neto2019scattering}. 
In short, given our experimental conditions, 
 the two alternative approaches amount to suppress or include
  the unscreened zero-frequency contribution in  (\ref{scatteringformula}). 
 
In the next section, we present our measurements and compare the results for the interaction energy and force with 
the theoretical models discussed above. 

\section{\label{sec:III}Interaction measurements and comparison with theoretical models}
\begin{figure}[h]
\includegraphics[scale=0.32]{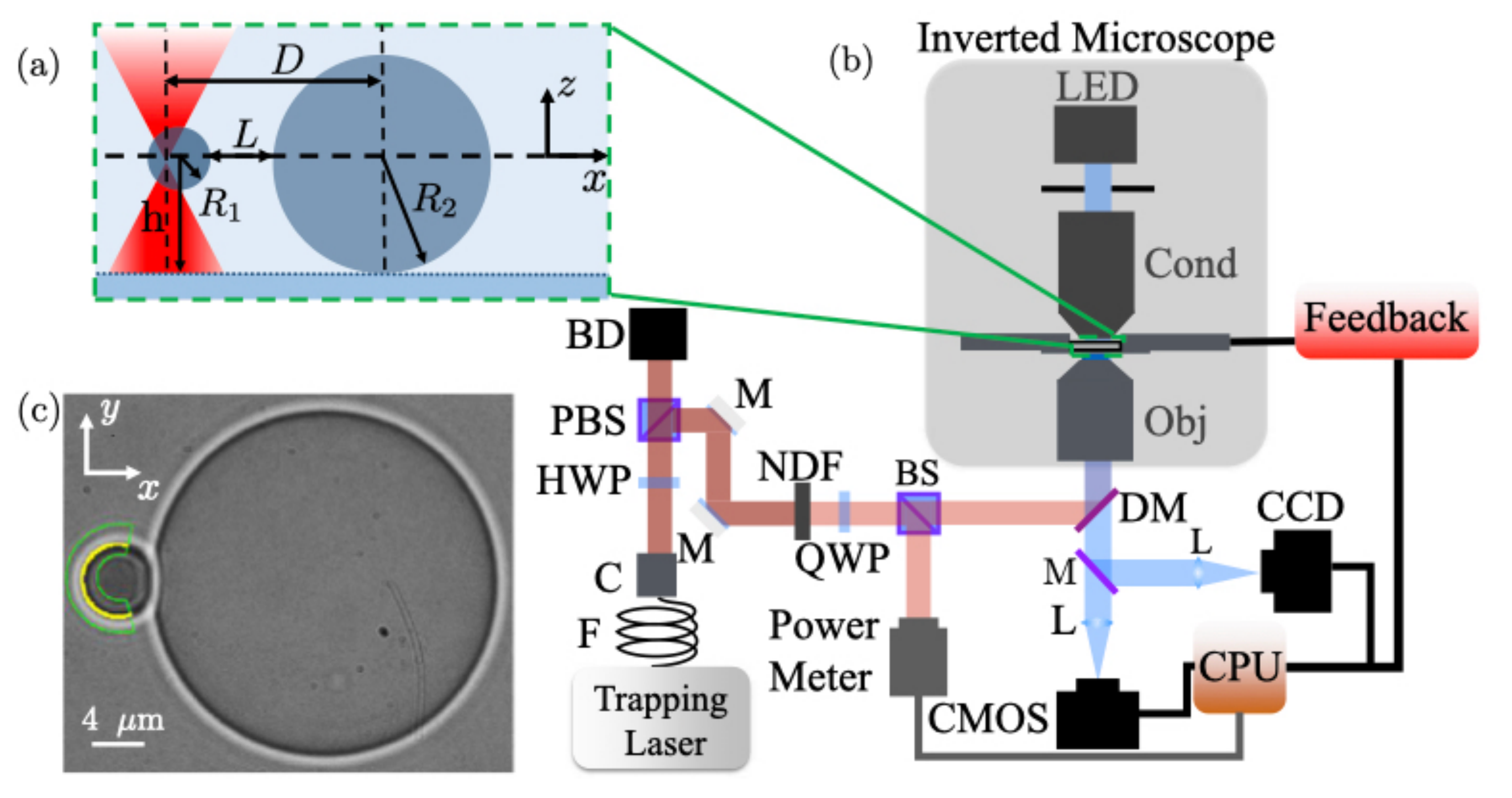}
\centering
\caption{(a) Sketch of two interacting silica microspheres separated by a distance $L.$
 We measure the Brownian fluctuations of the smaller one (radius $R_1$), which is optically trapped at a height $h$ from the coverslip. 
The larger microsphere (radius $R_2$) is adhered to the coverslip at a distance $D$ from the laser beam axis. 
(b) Experimental scheme for trapping, stabilization and position measurement:
 (F) optical fiber; (HWP) half-wave plate; (PBS) polarising beam splitter; (BD) beam dump; (M) dielectric mirror; (NDF) neutral density filter; (QWP) quarter-wave plate; (BS) balanced beam splitter; (DM) dichroic mirror; (L) lens; (Obj) water-immersion objective lens; (Cond) condenser lens; (CMOS) complementary metal-oxide-semiconductor camera; (CCD) charged-coupled device camera; (CPU) central processing unit.  (c) A typical optical image captured by the CMOS camera. In this case, $L$ corresponds to a few hundred nanometers. The green contour indicates
 the area used to detect the position of the trapped microsphere. The yellow dots are used to fit a circumference whose center represents the position. The scale 
 bar corresponds to $4\,\mu{\rm m}.$}
\label{fig:1}
\end{figure}

We employ a standard optical tweezers setup to probe the 
colloidal interactions between two silica microspheres of different radii.  Figure \ref{fig:1} presents a sketch of the experimental setup.
While the larger microsphere is adhered to the coverslip, the smaller one is optically trapped, as illustrated by panel (a). 
In order to align the two microsphere centers along the  $z$ axis with a precision $\delta z \sim 100\,{\rm nm},$ we follow~\cite{ether15} and use the information from
defocusing microscopy \cite{Agero2003,Gomez2021} to drive the microscope stage with our nano-positioning system.
The experimental scheme is shown in panel (b), while 
panel (c) presents a typical optical image of the two interacting microspheres.
 Details regarding the experimental setup, sample preparation and characterization of the microspheres can be found in Appendices \ref{sec:system},  \ref{sec:sample}
and  \ref{sec:characterization}, respectively.

The interaction is carried out under two conditions corresponding to very different values of the Debye screening length $\lambda_{\rm D}$.
A specific experimental protocol is implemented to each of these two conditions, as detailed in the next two subsections. 
We first present the methodologies which are common to both situations. 
 We measure the position of both microspheres by applying an edge detection method~\cite{ueberschar2012novel, yucel2017new}, as outlined in Fig. \ref{fig:1}(c) for the smaller microsphere (see appendix \ref{sec:detection} for details).
The resulting time series
 $(X_{1}(t),Y_{1}(t))$ and $(X_{2}(t),Y_{2}(t))$ for the smaller and larger microsphere, respectively, are the main ingredients for the analysis leading to the experimental energy and force data. 

We choose a very soft transverse trap stiffness $k_x\sim 1\,\si{fN/nm}$ so as to allow for femtonewton force measurements given the nanometric precision of our position detection method (see Appendix~\ref{sec:detection}). 
A major obstacle limiting the minimum distance that we can probe is the sharp increase of the Brownian 
correlation time  $\tau_{\rm C}=\gamma/k_x$ ($\gamma=$ Stokes drag coefficient) 
as the distance $L$ between the microspheres decreases for a fixed stiffness. 
Indeed, a longer measurement time would be needed to obtain the same amount of statistically independent values of  $(X_{1}(t),Y_{1}(t))$ when 
probing at shorter distances.
From the analysis of the position fluctuations, we measure the correlation time of the trapped particle under typical experimental conditions to be  
$\tau_{\rm C}\sim 100~\si{ms}$ 
for an average distance $\overline{L} \sim 1~\si{\mu m}.$ This is 
in agreement with the theoretical modelling of the drag coefficient $\gamma$ in the sphere-sphere geometry~\cite{cooley1969slow} when taking into account the Fax\'en 
correction~\cite{Faxen1922} arising from the coverslip for an average height $\overline{h}\sim 12~\si{\mu m}.$
We predict a fourfold increase of  $\tau_{\rm C}$ 
as the distance is reduced down to the minimum value
 $L\sim 0.2\,\mu{\rm m}$ probed in our experiment due to the variation of the drag coefficient $\gamma$~\cite{cooley1969slow}.
At distances shorter than this minimum value, the Brownian motion becomes too slow,
 making it impractical to probe the potential landscape.

To overcome the challenge imposed by the large correlation time,
we measure the positions over a very long time, $\mathcal{T}=500~\si{s}.$
Such a long measurement time requires mitigation of environmental noises and drifts so as to have a thermally-limited system 
over a time of the order of  $\mathcal{T}$ or longer.  
In Appendix \ref{sec:env}, we discuss the methodologies employed to render our 
system more stable over longer times, and how we have characterized and tested its stability. In particular, we 
show that our system is thermally-limited at least up to a time $\approx 2\mathcal{T},$ by analyzing the Allan deviation of the position variable $X_1.$

In view of the magnitude of the correlation time, we choose a sampling interval $1/f_s=100\,\si{ms}$ ($f_s=$ sampling rate)
to avoid storing correlated data, 
although the data becomes increasingly more correlated as the distance decreases. 
In addition, we take a short exposure time $W=2~\si{ms}\ll \tau_{\rm C}$  to avoid the blur effect arising from Brownian fluctuations while an individual frame is captured~\cite{wong06}.
The latter allows us to take the raw data for the position time series without introducing any model-dependent correction. 
 
For each experimental run, $5000$ frames are converted into the time series $(X_{1}(t),Y_{1}(t))$ and $(X_{2}(t),Y_{2}(t))$ while a feedback stabilization loop, shown in Fig.~\ref{fig:1} and discussed in detail in Appendix~\ref{sec:stage_drift}, keeps the microscope stage at a fixed position within $10\,\si{nm},$ 
which is the range of variation of $(X_{2}(t),Y_{2}(t)).$ 
Discrete binning is then implemented for the position of the optically trapped microsphere, with a bin size of $4~\si{nm}$ comparable to our position resolution. 
As indicated by Fig.~\ref{fig:1}(a), 
the distance between the microspheres for a given bin $i$ is given by
\begin{equation}\label{Li}
L^{(i)} = \overline{X}_{2}-X^{ (i)}_{1} - (R_{1} + R_{2}),
\end{equation}
where $\overline{X}_{2}$ corresponds to the average position of the adhered microsphere and 
$X^{(i)}_{1}$ is the position of the optically trapped one. 
The microsphere radius  $R_{1}=(2.35\pm 0.02)~\si{\mu m}$ is measured by scanning electron microscopy (SEM),
with the error given by the standard deviation (see Appendix~\ref{sec:sem}).
Since our batch of the large silica microspheres presented an important size dispersion, we determined 
their radii from the correlation of optical and SEM images, as discussed in Appendix~\ref{sec:corrmic}.
We found
$R_{2} = \left( 12.76 \pm 0.06 \right)~\si{\mu m}$ and 
$R_{2} = \left( 11.74 \pm 0.06\right) ~\si{\mu m}$
 for the microspheres used with low (Sec.~\ref{sec:III}A) and high  (Sec.~\ref{sec:III}B) salt concentrations, respectively. While the error of 
 the center-to-center distance $\overline{X}_{2}-X^{ (i)}_{1}$ is in the nanometer range, 
 the global offset arising from the subtraction of $R_1+R_2$ has a higher error which is not 
written explicitly as it 
 does not change the relative values of $L^{ (i)}.$

The binned data for each experimental run lead to a frequency histogram from which we obtain 
the probability distribution $p (L^{ (i)}).$
Since our system remains in thermal equilibrium over a time scale longer than the total measurement time ${\cal T},$
 we associate $p (L^{ (i)})$ to a Boltzmann distribution~\cite{florin1998,paladugu2016nonadditivity}
 at the measured temperature $T=(296 \pm 1)\,\si{K}.$ 
 For each salt concentration, we choose a reference distance $L_{{\rm ref}}$
 with a high number of occurrences for all different runs 
so as to be able to accurately determine $p(L_{{\rm ref}})$ from the available data.  
We then infer the total potential energy
$U (L^{ (i)})$  from 
\begin{equation}
U (L^{ (i)})-U(L_{{\rm ref}})= -k_{B}T\left[\log p(L^{ (i)}) - \log p(L_{{\rm ref}}) \right]\, , 
\label{eq.prob2}
 \end{equation}
where $k_B$ is the Boltzmann constant.

Eq.~(\ref{eq.prob2}) determines the potential energy from the probability distribution $p (L^{ (i)})$
apart from
an arbitrary offset as expected on physical grounds.
After subtracting the optical potential $U_{\rm opt}(L^{(i)})$, which, as discussed below, is determined by measurements when the two microspheres are sufficiently apart,
we average
the results for the interaction energy 
 $U_{\rm int}(L^{(i)})=U (L^{ (i)})-U_{\rm opt}(L^{ (i)})$ 
from different runs. 
From now on we will omit the bin index~$i$. 
 
 In addition to the environmental noise discussed in detail in Appendix~\ref{sec:env}, one important concern is the modification of the optical force as the 
adhered microsphere is brought closer to the laser focal spot. 
We expect the reverberation of the laser beam between the interacting microspheres to be negligible since their refractive index $n_{\rm bead}=1.4146$ is close 
 to the refractive index of the host medium $n_{\rm water}=1.3242$ (see Appendix~\ref{sec:cal} for details).  Indeed, the corresponding Fresnel reflectivity is as small as $0.1\%$ for normal incidence. 
 In addition, the optical reverberation between the trapped particle and the coverslip \cite{dutra2016theory} is unimportant in our experiment as the former is kept at a fixed height with respect to the latter. 
 
Yet optical perturbations might still be present as part of the trapping beam is refracted through the adhered microsphere before reaching the focal plane.
 In contrast to the reverberation effect, such perturbation does not involve interference and is thus expected to 
 depend only on the relative refractive index and the geometrical aspect ratio $D/R_2,$ where $D$ is the distance between the (unperturbed) focal point lying along the optical symmetry axis 
 and the center of the adhered sphere as indicated in Fig.~\ref{fig:1}(a). 
 Preliminary theoretical results~\cite{Schoger2020} indicate a total displacement of the optical equilibrium position along the $x$ axis of a few tens of nanometers  
 and a variation of stiffness of a few percent as the adhered sphere is brought from infinity to 
 a distance of closest approach of a few hundred nanometers.  

In view of the above results, we measure the optical equilibrium position $X_{1,{\rm eq}}^{\rm opt}$ and the 
trap stiffnesses $k_x$ and $k_y$ (see Appendix \ref{sec:stage_drift}), characterizing the optical potential $U_{\rm opt}(L)$,
at intermediate distances $D$ such that the
surface-to-surface distance at optical equilibrium 
 $L^{\rm opt}_{\rm eq}$ lies in the interval from $480\,\si{nm}$ to $800\,\si{nm}.$
In this range, colloidal interactions in our system are negligible and then $X_{1,{\rm eq}}^{\rm opt}$ is obtained from the average position.
More importantly, our calibration of the optical equilibrium position already takes the beam perturbation into account. 
As the geometrical  aspect ratio $D/R_2$ changes by only $\sim 2.5\%$ within this interval, we expect 
the optical potential to be approximately independent of $D.$ 
We have verified, experimentally, that $k_x,$ $k_y$ and  $X_{1,{\rm eq}}^{\rm opt}$
(see Appendix~\ref{sec:drift}) are indeed
independent of $D$ within the experimental error in this range of distances. 

In order to push further into the interaction region, we reduce the distance $D$ between the adhered sphere and the 
optical symmetry axis by up to $\sim 150\,{\rm nm}$
with respect to the smallest value of $D$ employed when characterizing the optical potential. 
We assume that $X_{1,{\rm eq}}^{\rm opt}$ and $k_x$ are still constant in this case.
This is verified by measuring $k_y$ (see Appendix \ref{sec:cal}),
while the total stiffness $k_x-\partial_x F_{\rm int}$ and the equilibrium position change due to the colloidal interactions.
To further verify that our interaction measurements are not contaminated by perturbations of the optical force, 
we perform measurements at different laser powers, as discussed in the next subsection.

\subsection{\label{sec:low}Low salt concentration}
 
In this subsection we present the results for the colloidal interaction 
and discuss some additional experimental details which are specific to the experiment with  no added salt. 
In this case, the double-layer force is dominant and the total force is repulsive over the entire probed range of distances.
In order to overcome the repulsive interaction and 
adhere the larger microsphere to the coverslip as indicated in Fig.~\ref{fig:1}(a), we coated the
latter with poly-L-lysine (see Appendix \ref{sec:sem}). 

We performed 25 interaction runs for the same pair of microspheres,
each run corresponding to fixed values for the distance $D$ and the laser power.
We employed three different values for $D$ and 
several values for the laser power,
 with the resulting $k_x$ ranging from $1.2~\textrm{to} ~2.5\, \si{fN/nm}$
(see Fig.~\ref{fig:drift}(a) of Appendix~\ref{sec:env} for details).
  In addition, we performed four calibration measurements,
 employing a larger distance $D,$ in order to determine the optical equilibrium position
 $X_{1,{\rm eq}}^{\rm opt}$,  which
 was such as to correspond to a surface-to-surface separation $L^{\rm opt}_{\rm eq}=480\,\si{nm}.$
After subtracting the optical potential $U_{\rm opt}(L)$
from the total potential energy $U(L),$
 we find no systematic effect 
of the laser power on the measured interaction energy,
as the data for $U_{\rm int}(L)$ coming from different runs are compatible with each other and do not split according to the laser power.
Indeed, any additional perturbation of the optical potential as the adhered microsphere is brought closer to the trapped one would result 
in the contamination of  $U_{\rm int}(L)$ by a residual optical potential proportional to the laser power after subtraction of the unperturbed optical potential.
Thus, our analysis indicates that such additional perturbation is
negligible when compared to our experimental sensitivity.

\begin{figure}[h]
\includegraphics[scale=0.24]{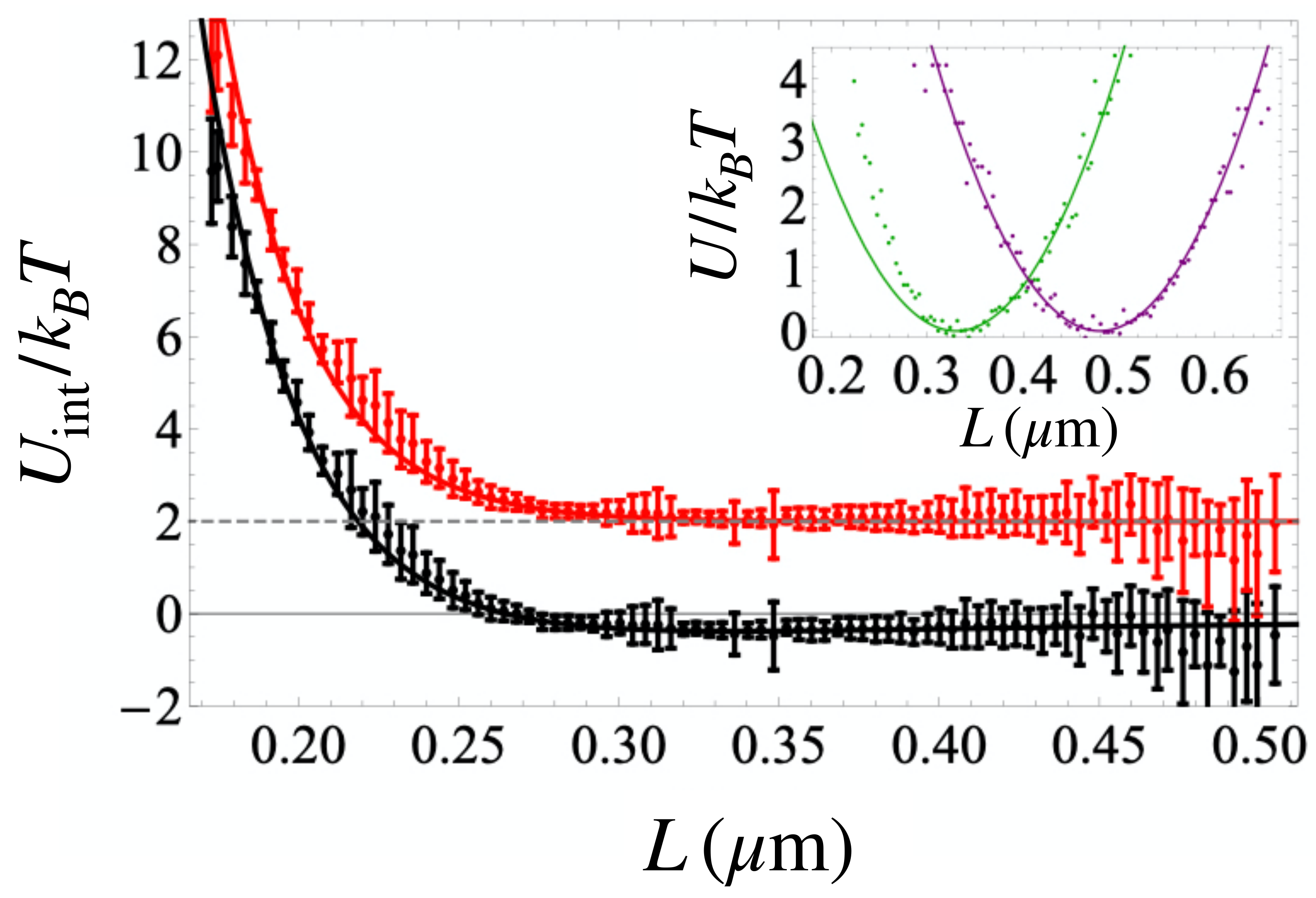}
\centering
\caption{Interaction energy in units of $k_B T$ versus distance for the sample with a low salt concentration:
experiment (points with error bars representing the standard error of the mean), and theoretical fits based on
two different theoretical models of the Casimir interaction, either with (black) or without (red) the unscreened contribution from TM modes in the limit of zero frequency. 
The latter is displaced by $ 2 \,k_{B} T $ and the same experimental data is plotted twice for better visualization.
For both models, 
the double-layer repulsive interaction is calculated within the linear superposition approximation in terms of two fitting parameters: the squared charge density $\sigma^2$ and the Debye screening length $\lambda_{\rm D}.$
In the inset, we plot the total energy versus distance for two individual runs.
 The purple and green plots 
correspond to the adhered microsphere placed further and closer to the laser beam axis, respectively. In both plots, the solid line represents the optical potential. 
 }
\label{fig:2}
\end{figure}

We average the data for $U_{\rm int}(L)$ from the 25 runs, combining different laser powers and values of $D$, and plot the results in Fig.~\ref{fig:2}.
The error bars represent the standard error of the mean of the combined set of data and the common reference distance is
 $L_{{\rm ref}} = 0.27~\si{\mu m}$ \cite{foot}.
For the purpose of illustrating how those results are obtained, we also
plot the total energy $U(L)$ corresponding to two individual runs
in the inset. 
While the purple points correspond to the calibration configuration used to characterize $U_{\rm opt}(L)$, the green ones represent a single interaction run with a smaller $D$ and the same laser power taken here as a typical example. 
We determine the optical equilibrium position  $X_{1,{\rm eq}}^{\rm opt}$ as the position average given the probability distribution corresponding to the purple points (no interaction). 
 The surface-to-surface optical equilibrium position  $L^{\rm opt}_{\rm eq}=480\,{\rm nm}$ is then obtained from
  $X_{1,{\rm eq}}^{\rm opt}$ by using Eq.~(\ref{Li}). The optical stiffness $k_x$ is derived from the experimental calibration methods described in Appendix~\ref{sec:cal}.
The resulting optical potential $U_{\rm opt}(L)=k_{x}(L-L_{\textrm{eq}}^{\textrm{opt}})^{2}/2$ is represented by the solid purple curve, which is also a good quadratic fit of the purple points as expected. 
Finally, the green solid curve is the optical potential with the same stiffness and $X_{1,{\rm eq}}^{\rm opt}$ but with a different value of  $L^{\rm opt}_{\rm eq}$ as determined by
Eq.~(\ref{Li}), since it corresponds to a smaller value $\bar{X}_2$ for the position of the adhered microsphere. 
The resulting displaced curve represents the optical potential $U_{\rm opt}(L)$ in the interaction run taken as example. 
When compared to the experimental curve for $U(L)$ (green points), it makes more apparent the skewness of the latter, which indicates the repulsive nature
of the colloidal interaction. 
Finally, the data for $U_{\rm int}(L)$ coming from this specific run is then the difference between the green points and the green solid curve shown in the inset.

In order to allow for comparison with two distinct theoretical models of Casimir screening, 
we plot in Fig.~\ref{fig:2} the same experimental results for $U_{\rm int}(L)$ twice for better visualization,
with the red points shifted by $ 2\, k_{B} T$ with respect to the black ones. 
The curve fits are then based on two distinct models of the Casimir attractive interaction, either
with (black) or without (red) the contribution arising from TM channels in the zero-frequency limit (see Sec.~\ref{sec:inte} for details).
In both cases, the double-layer repulsive interaction is calculated within the LSA and obtained from Eq.~(\ref{eq.dcLSA})
 in terms of two fitting parameters: the squared charge density $\sigma^2$ and the Debye screening length $\lambda_{\rm D}.$
We fit over the interval $L_{\rm min}\le L \le 440\,{\si{nm}}$ and take values for $L_{\rm min}$ between $190\,\si{nm}$ and $210\,\si{nm}.$
We also consider different data sets obtained by changing the reference distance  $L_{{\rm ref}}$ by a few tens of nanometers (not shown in the plot). 
In Table~\ref{tab.fit}, we show the average of the fitted values of $\sigma$ and $\lambda_{\rm D}$ with
the errors representing the standard deviation.

The red curve in Fig.~\ref{fig:2} corresponds to a repulsive colloidal potential over the entire range of distances probed in our experiment. 
On the other hand, the black solid line exhibits a very weak attraction for distances $L\gtrsim 0.35\,\mu{\rm m},$ since in this range the TM zero-frequency contribution is much larger than the combined contribution of all nonzero Matsubara frequencies. 
However, such smooth variation is below our experimental sensitivity. 
Thus, both models fit our data equally well, although
the fitting parameters are more stable with respect to the fitting range when the zero-frequency TM contribution is included, as
indicated by the standard deviations shown in Table~\ref{tab.fit}.

  \begin{table}[H] 
 \caption{Parameters employed for the curve fit of the measured interaction energy: charge density $\sigma$ and 
 Debye screening length $\lambda_{\rm D}.$ In addition to the double-layer interaction energy (\ref{eq.dcLSA}), we also
 consider the Casimir interaction either with or without the zero-frequency TM contribution. 
\label{tab.fit}}
  \begin{ruledtabular}
 \begin{tabular}{ccc}
Casimir model & $\sigma~\left( \si{mC/m^{2}}\right)  $ &  $\lambda_{\rm D}~ \left(\si{nm}\right) $ \\
\hline
 $n=1,2,...$   & $-1.7\pm 0.3$ & $25.0\pm 0.9$ \\
$n=0,1,2,...$ & $-0.8\pm 0.1$  & $29.3\pm0.6$ 
 \end{tabular}
 \end{ruledtabular}
 \end{table}
 
The results for $\lambda_{\rm D}$ shown in Table~\ref{tab.fit} differ by merely $\sim 20\%.$
 They correspond to a salt concentration $n_{\infty}\sim  0.2~\si{mM},$
 which is compatible with sample contaminations
 arising, for instance, 
from the poly-L-lysine coating employed to adhere the larger microsphere to the coverslip.
 The slightly larger value of $\lambda_{\rm D}$ obtained 
 when including the zero-frequency TM contribution 
 is required to yield a double-layer repulsion with a longer range that compensates for the stronger Casimir attraction in this case (see subsection~\ref{sec:high}). 
As indicated in Table~\ref{tab.fit},  
the corresponding  charge density $\sigma$ is then
lowered by a factor~$\sim\!2$ in order to also fit at shorter distances.
We conclude that the fitted value of $\sigma$ depends strongly on the theoretical model of the Casimir interaction even in the distance range $L>0.1\,\mu\si{m}$ where it is clearly sub-dominant. 
Model-dependent results for colloidal parameters were also reported in Ref.~\cite{nayeri2013measurements}.
 
 Usually, the surface charge density is obtained by fitting the interaction force at much shorter distances, $L \lesssim 20\,\si{nm},$
 by considering a Casimir (vdW) model 
 without the zero-frequency TM contribution. 
 Available results for the silica charge density for several symmetric inorganic salts at $\sim 0.2\, \si{mM}$ tend to cluster, for a $\si{pH}=5.6$, 
 at $\sigma \sim -4\,\si{mC/m^{2}}$~\cite{smith2020forces}.
 Since the magnitude of the silica charge density increases with an increasing pH~\cite{behrens01}, 
 the discrepancy with respect to 
 the values shown in Table~\ref{tab.fit} cannot be attributed to the higher pH of our sample ($\si{pH}=6.8$).
Theoretical modelling of silica charging~\cite{behrens01} seems to
 favor the values shown in Table~\ref{tab.fit}, and especially the one found when including the TM zero-frequency contribution. 
 
The two models of Casimir screening 
can hardly be distinguished in
experiments with low salt concentrations,
since the Casimir interaction is sub-dominant over the entire distance range probed experimentally. In order to isolate the Casimir interaction from 
the electrostatic double-layer signal, we performed an experiment with a much higher salt concentration as presented in the next sub-section. 
 
\subsection{\label{sec:high}High salt concentration}

When the double-layer interaction is totally suppressed by ionic screening,  the force signal at distances $L\gtrsim 0.2\,\mu\si{m}$
is considerably weaker, making it harder to measure. Moreover, the equilibrium position of the probe particle becomes more unstable, due to 
the sharp increase of the attractive Casimir force at short distances. 
Experimentally, it is difficult to control the distance $D$ between the adhered microsphere and the laser beam 
so as to hit the narrow range of distances in which the Casimir attraction is
 measurable and yet not strong enough to make the probe jump into contact in the beginning of the run. 

We prepare a sample with a NaI concentration of  $n_{\infty}= 0.22 ~\si{M},$
 corresponding to $\lambda_{\rm D} = 0.64 ~\si{nm}$ according to Eq.~(\ref{eq.lambdaD}).
Such value for the Debye screening length is sufficiently small to produce a complete suppression of the double-layer interaction for $L\gtrsim 0.2\,\mu\si{m}.$
The interaction runs are performed for average separations of the order of $ \overline{L} \sim 0.4\,\mu\si{m}.$ To determine the  parameters of the optical potential alone, 
we take 8 different average separations in the range $0.5 \,\mu\si{m}< \overline{L}< 0.9\,\mu\si{m}.$ Those calibration measurements are repeated before and after every two interaction measurements.
In all cases, we find variations of the optical equilibrium position and trap stiffnesses $k_x$ and $k_y$ comparable to the corresponding experimental errors.
The laser drift is negligible during each run and also from one run to the next, as discussed in Appendix \ref{sec:drift}. 
In order to combine all interaction runs, we take $L_{\rm ref} = 455~\si{nm}$ as the reference separation. All measurements are performed with the same laser power. 

\begin{figure}[h]
  \includegraphics[scale=0.21]{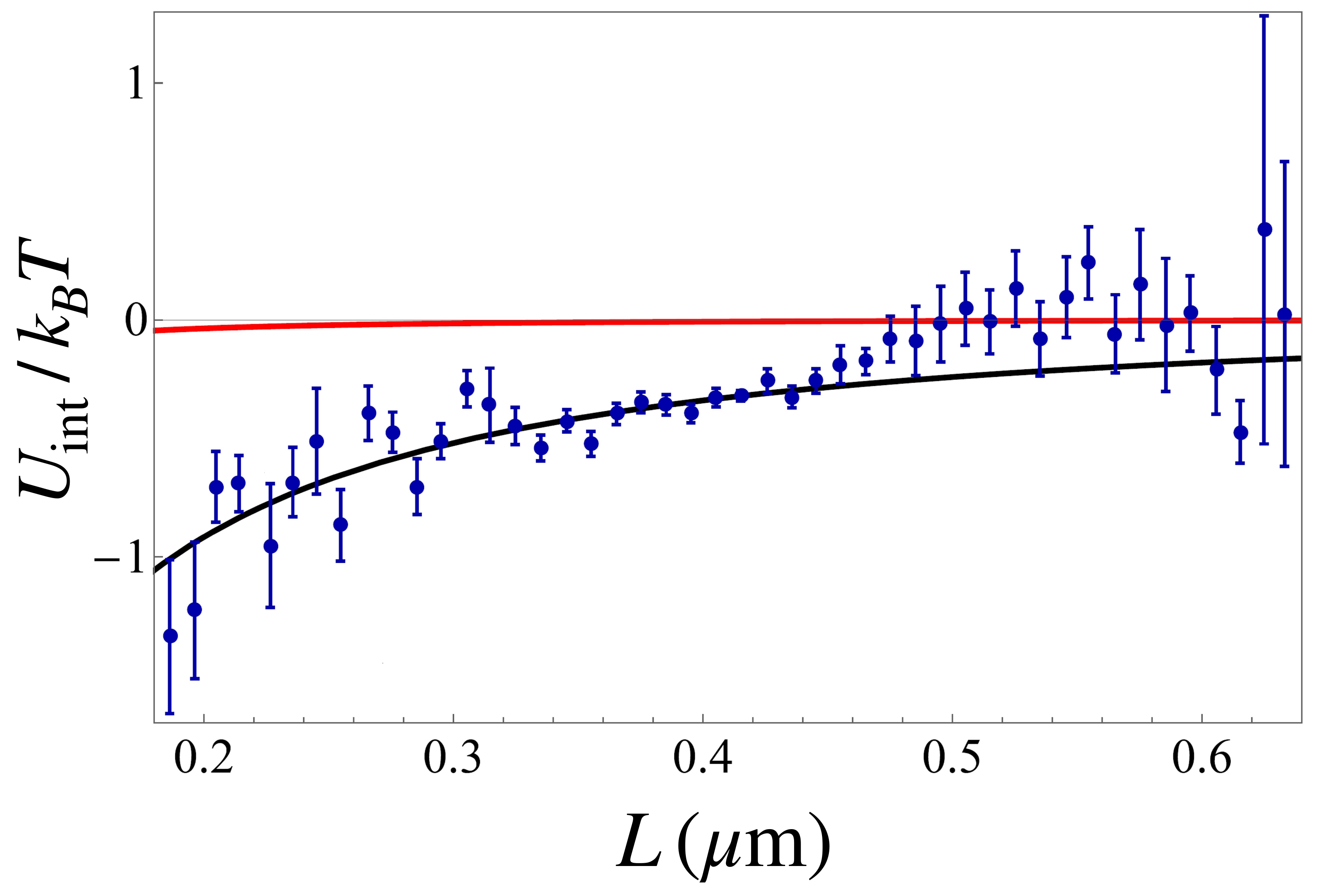}
\centering
\caption{Casimir energy in units of $k_B T$ versus distance for the sample with a high salt concentration: experiment (points with error bars representing the standard error of the mean)
and theory either 
 with (black) or without (red) the unscreened contribution from TM modes in the limit of zero frequency. 
The salt concentration of $n_{\infty}= 0.22 ~\si{M},$
corresponding to a screening length 
 $\lambda_{\rm D}=0.64$ nm, is such that the double-layer interaction is completely suppressed in the distance range probed experimentally. }
\label{fig:3}
\end{figure}

We subtract the optical potential from the total potential and
plot the resulting Casimir energy versus distance in Fig.~\ref{fig:3}.
 We compare the experimental data (points) with the two different screening models for the Casimir interaction: 
 with (black solid line) and without (red solid line) the 
 unscreened contribution from TM modes in the limit of zero frequency.
 No fitting procedure is implemented here, as the theoretical Casimir energy is obtained exclusively from the known dielectric functions of silica and water \cite{Zwol2010}.
Because of the near index-matching between silica and water at nonzero Matsubara frequencies, the 
standard procedure \cite{parsegian2005van} of disregarding TM scattering channels in the zero-frequency limit leads to a negligibly small interaction energy, which cannot explain 
our data as indicated in Fig.~\ref{fig:3}, even if
one takes the distance offset as a fitting parameter.
On the other hand, a blind comparison with the full scattering theory shows a good agreement with the experimental data.

In addition to the interaction energy, we also probe the  interaction force 
$F_{\rm int}=k_x\, (L_{\rm eq}^{\rm int}-L_{\rm eq}^{\rm opt}),$
with negative sign denoting attraction,
by measuring the equilibrium distance  $L_{\rm eq}^{\rm int}.$ 
The equilibrium distance does not necessarily coincide with the average distance $\bar{L}$ 
because the probability distribution  $p (L)$ has a negative skewness 
that results from the nonlinear Casimir attraction. 
More specifically, using a quadratic polynomial to fit the potential $U(L)$  overestimates 
the magnitude of the force as  $\bar{L}<L_{\rm eq}^{\rm int}$ in the case of negative skew.
Thus, we determine  $L_{\rm eq}^{\rm int}$ by fitting the potential with a cubic polynomial when 
considering experimental runs corresponding to asymmetric probability distributions. 
On the other hand, using the cubic polynomial for the runs with the largest values of $D$ leads to overfitting.
Indeed, the potential is expected to be approximately quadratic as the interaction is negligible when the adhered microsphere is placed far from the laser beam.
We use the absolute value of the skewness $|\mu_3|\equiv |\overline{[(L^{(i)}-\bar{L})/\sigma_x]^3}|$ ($\sigma_x=$ standard deviation)
of the probability distribution for each experimental run (corresponding to a fixed distance $D$) 
to define which polynomial degree is taken in the fitting procedure: 
quadratic for $|\mu_3|\le 0.05,$ and cubic otherwise.
 All measured values of $\mu_3$ are compatible with theoretical predictions for the 
 second derivative of the Casimir force provided that the TM zero-frequency contribution is included.
 
\begin{figure}[h]
\includegraphics[scale=0.21]{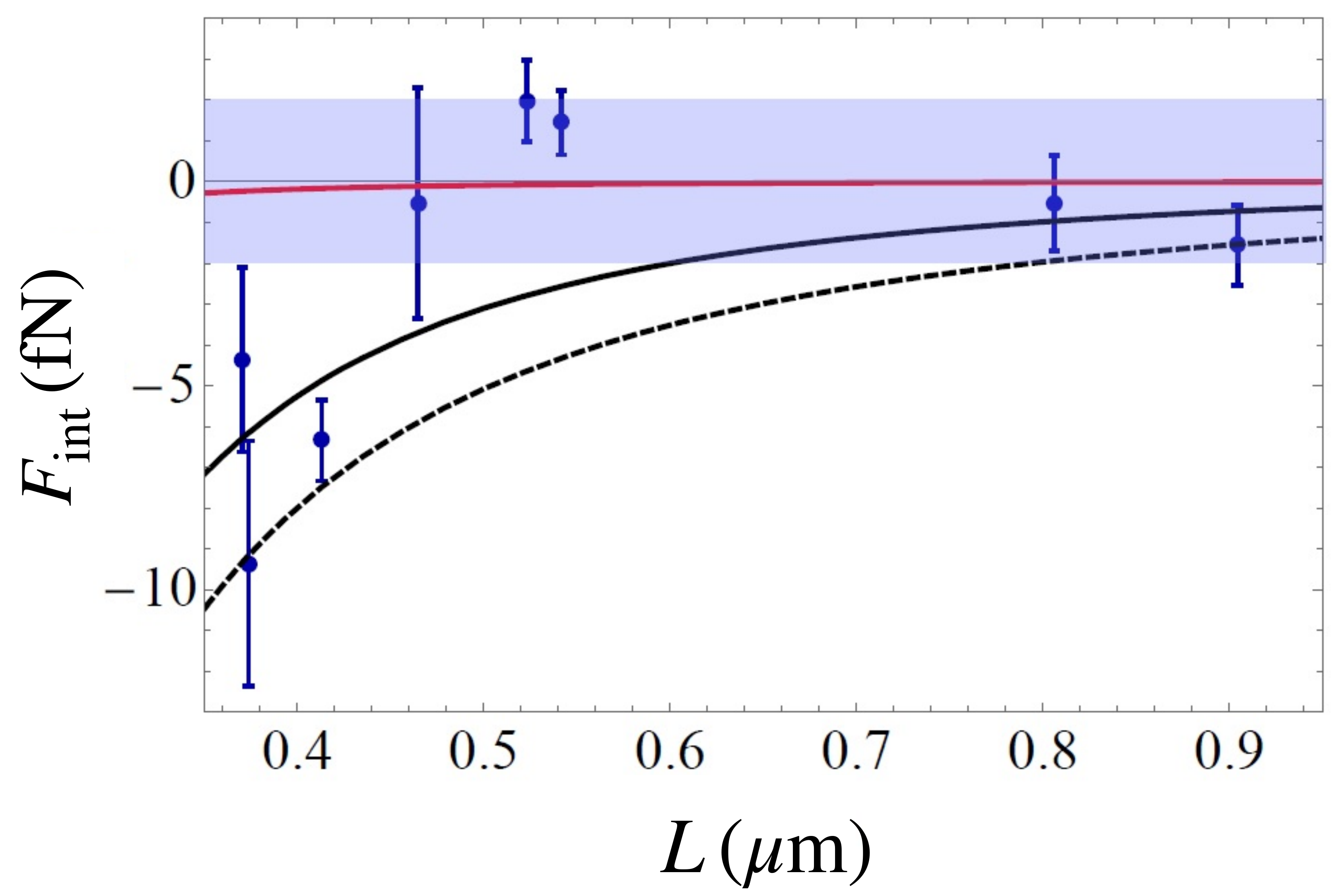}
\centering
\caption{ Casimir force versus distance: experiment (points with error bars) and theory either 
 with (black) or without (red) the unscreened contribution from TM modes in the limit of zero frequency. 
 The dashed curve corresponds to the proximity force approximation with 
TM zero-frequency modes included. 
  The light blue band indicates the experimental sensitivity for force measurements in our setup. 
 }
\label{fig:4}
\end{figure}

In Fig.~\ref{fig:4}, we plot the interaction force as a function of the separation distance $L=L_{\rm eq}^{\rm int}$ in equilibrium. In contrast with the 
potential energy plots of Figs.~\ref{fig:2} and \ref{fig:3}, here each data point is the result of a fit of the entire potential obtained from a given run 
corresponding to a fixed value of $D.$
 The shaded area indicates the sensitivity $2\,\si{fN}$ of our force measurement, which is determined by the error of the equilibrium position
as discussed in Appendix \ref{sec:drift}. 
As in the case of the Casimir potential shown in Fig.~\ref{fig:3}, we find agreement only with 
the theoretical prediction including the contribution of TM channels in the zero-frequency limit (solid black line). 
The contribution of nonzero Matsubara frequencies (red) is smaller by about one order of magnitude and cannot describe our data. 
Previous experiments under similar conditions~\cite{hansen2005novel,kundu2019measurement} also found a 
signal larger than predicted by the standard theoretical model that excludes TM modes in the zero-frequency limit. 
 
We also plot the theoretical results (dashed) obtained by considering the spherical geometry within PFA~\cite{Derjaguin1934}, with the 
TM zero-frequency contribution included. The PFA provides a direct connection between the spherical geometry and the parallel-planes one considered in Ref.~\cite{neto2019scattering} 
and is asymptotically valid for large aspect ratios $R_{\rm eff}/L\gg 1$ \cite{Spreng2018}.
Although PFA overestimates the exact Mie scattering results for the force by $\sim 50\%$ for the parameters corresponding to Fig.~\ref{fig:4},
our data do not allow for a discrimination between 
 the two theoretical models that include the TM channels in the zero-frequency limit.

\section{\label{sec:conc}Conclusion}

We developed a protocol for using optical tweezers to measure the surface interaction 
between silica microspheres  separated by distances above $0.2\,\mu{\rm m}$ in aqueous solution.
For the sample with the highest salt concentration, 
the measured potential energy corresponds to an attractive Casimir force in the 
femtonewton range.  We find good agreement (no fitting) between our experimental data and the 
scattering theory which contains the unscreened contribution of transverse magnetic modes in the zero-frequency limit. 
 Such contribution dominates the total Casimir signal by roughly one order of magnitude and is not included 
in the standard description of the van der Waals interaction across ionic solutions.   

 When measuring the surface interaction at lower salt concentrations, the theoretical description of the Casimir effect has an impact on the characterization of the repulsive double-layer interaction, particularly on the fitted value for the surface charge density. 
 
The higher salt concentration employed in the Casimir experiment is comparable to typical values found in living cells~\cite{Liu2017}. 
Thus, the indication that the Casimir interaction is stronger and of a longer range under such conditions might have important implications in the fields of 
cell and molecular biology. 

\begin{acknowledgments}

We thank Tanja Schoger for sharing preliminary theoretical results on the perturbation of the optical force by the presence of an additional microsphere.
We also thank Antoine Canaguier-Durand, Sergio Ciliberto, 
Anne Le Cunuder, Romain Gu\'erout, Astrid Lambrecht, Umar Mohideen, Jeremy Munday, George Palasantzas, Rudolf Podgornik, and Vitaly Svetovoy for 
enlightening discussions.
 We acknowledge funding from the Brazilian agencies 
 Conselho Nacional de Desenvolvimento Cient\'{\i}fico e Tecnol\'ogico (CNPq--Brazil),
Coordena\c c\~ao de Aperfei\c coamento de Pessoal de N\'{\i}vel Superior (CAPES--Brazil),  Instituto Nacional de Ci\^encia e Tecnologia Fluidos Complexos  (INCT-FCx), and the
Research Foundations of the States of Minas Gerais
(FAPEMIG), Rio de Janeiro (FAPERJ) and S\~ao Paulo
(FAPESP). 
This work was also supported by Centre National de la Recherche Scientifique (CNRS),
Sorbonne Universit\'e, and Deutscher Akademischer Austauschdienst (DAAD)
through their collaboration programs 
 Projet International de Coop\'eration Scientifique (PICS), Convergence International, and PROBRAL, respectively.
We are grateful to the Instituto Nacional de Metrologia, Qualidade e Tecnologia (Inmetro)
for letting us use their high-resolution scanning electron microscope.

\end{acknowledgments}

\appendix

\section{\label{sec:system} Experimental Setup}
\par A schematic of our experiment is shown in Fig.~\ref{fig:1}(b) of the main text. A $1064~\si{nm} $ laser beam (YLR-5-1064LP, IPG Photonics) exits an optical fiber and is divided using a half wave plate and a polarized beam splitter. While the transmitted laser light is blocked, the reflected one is directed towards alignment mirrors, an attenuation neutral density filter, and a quarter-wave plate, fitted to change the laser beam well-defined linear polarization into circular polarization. In fact, this procedure is important in order to produce equal optical trap stiffnesses along the $x$ and $y$ axes in the plane orthogonal to the beam propagation axis $z$. The laser beam is then divided again by a balanced nonpolarized beam splitter, which halves the light into a power meter (1936C and 918D-UV-OD3R, Newport) and towards the microscope (Eclipse Ti-S, Nikon). Under the microscope, the laser beam hits a dichroic mirror (ZT532rdc-NIR-R725-1100-UF2, Chroma) with high reflectance for $ 1064~\si{nm}$ and high transmittance for visible light. 

A $60\times$ water-immersion objective (CFI60 Plan Apochromat VC, Nikon) with a numerical aperture of $ {\rm NA} = 1.2 $ focuses the laser light on the sample. 
The objective corrects the 
spherical aberration introduced by the index mismatch at the glass-water interface, then allowing reliable optical trapping and imaging at heights of dozens of micrometers above the  coverslip. In comparison with oil-immersion objectives, 
water-immersion  also requires shorter stabilization times, as water is less viscous than common immersion oils, consequently decreasing systematic errors associated with image drifts.

For brightfield illumination, we used a $470~\si{nm}$ wavelength light produced by a high-power mounted LED (M470L3, Thorlabs), chosen to also avoid heating effects in the sample. After passing through a ${\rm NA} = 0.85$ condenser lens, the light traverses the sample and is gathered by the objective lens. Transmitted afterwards by the dichroic mirror, the light is then recollected by the tube lens and directed into a CMOS (Orca-flash 2.8, C11440-10C, Hamamatsu) camera for measurements and visualization, and into a CCD camera for active feedback stabilization. The displacements in the sample are controlled by a digital piezoelectric controller (E710, Physik Instrumente) connected to the microscope stage, in which the sample chamber is attached. All the optical elements are mounted on an optical breadboard (M-SG-30x60-4, Newport), placed on a conventional optical table (RS-2000 Newport).

\section{\label{sec:sample} Sample Preparation }
\par Our sample is composed of a dispersion of two sets of uncoated silica microspheres with nominal manufactured radii of $\bar{R}_{1}=2.5~\si{\mu m}$ (Cat\#24332, Polysciences, Inc.) and $\bar{R}_{2}=10~\si{\mu m}$ (DNG-B020, DiagNano\texttrademark) diluted in an ultra-pure water solution (Millipore). A 1:1 salt (Sodium Iodate, Scientific NaI-Exodus) is added in the case of high salt concentration measurements. The prepared solution is then introduced inside the sample chamber composed of a rubber O-ring between two cleaned coverslips ($24~\si{mm}\times 60~\si{mm}$ and $ 24~\si{mm}\times 32~\si{mm}$, Knittel Glass) and properly sealed with silicone grease. For the calibration runs, the sample temperature is monitored by a thermocouple (5TC-TT-T-30-36, Omega) immersed in solution and connected to a thermometer (DP24-T, Omega). Over the range of laser powers ($\sim\si{mW}$) used in the interaction and calibration experiments, the temperature is given by $T=(296\pm1)\,\si{K}$ and no changes are observed.

\section{\label{sec:characterization}Characterisation of Microspheres}

\subsection{\label{sec:sem}Scanning Electron Microscopy (SEM)}

Knowing the radii of the microspheres is crucial to determine the distance offset from the relation  
$L = \overline{X}_2-X_1 - (R_{1} + R_{2})$  (see Figs.~\ref{fig:1}(a) and \ref{fig:1}(c)).
To perform this task, high resolution scanning electron microscopy (SEM) is applied on test samples containing separately one of the two sets of silica microspheres. A glass coverslip (Paul Marienfeld GmbH \& Co. KG, Germany) previously coated with 0.01\% poly-L-lysine (Sigma-Aldrich, Darmstadt, Germany) and containing the targeted microspheres is firstly fixed upon a metallic SEM stub using a conductive carbon tape (Pelco TabsTM, Stansted, Essex, UK). The glass coverslip is then dried with a weak jet of nitrogen gas to avoid contamination, and afterwards coated with a $4-5~\si{nm}$ thick platinum layer using a sputtering device (Leica EM SCD 500, Wetzlar, Germany). Finally, the sample is characterized using a scanning electron microscope (Quanta 450TM FEG, FEI Company, USA) operating at $ 5~\si{kV}$.
\par
\begin{figure}[h]
\includegraphics[scale=0.37]{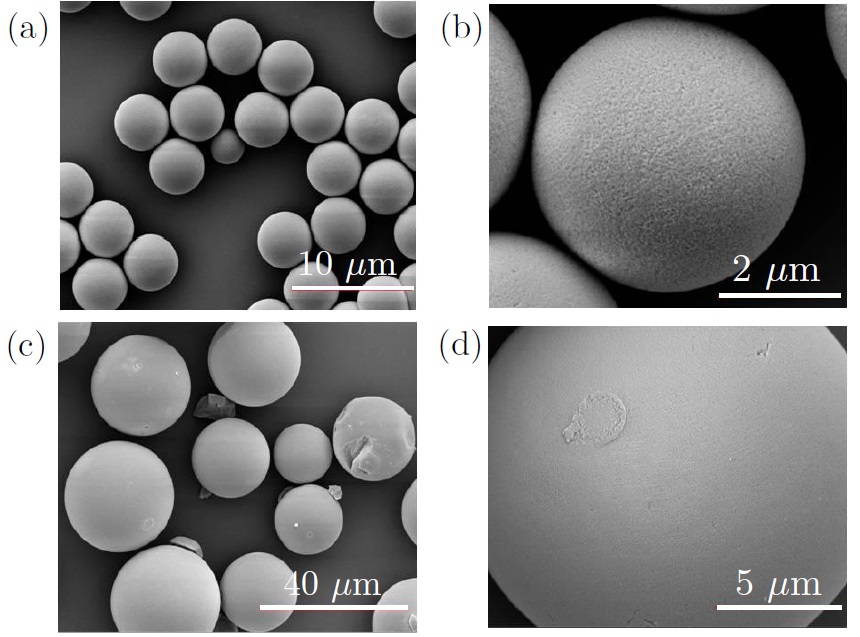}
\centering
\caption{Typical SEM images for
characterization of the silica microspheres employed in our measurements. Panels (a-b) and (c-d) correspond to batches
 with nominal radii $\bar{R}_{1}=2.5~\si{\mu m}$ and  $\bar{R}_{2}=10~\si{\mu m},$ respectively. 
 Scale bars are shown in each panel.}
\label{fig:5}
\end{figure}

 Typical SEM images are shown in Fig.~\ref{fig:5}. Panels  (a) and (b) correspond to the set of small silica microspheres, while (c) and (d) show images from the set of large ones. Figure~\ref{fig:5}(a) shows that the first group exhibits a small size dispersion, with radius of $R_{1}=(2.35\pm 0.02)~\si{\mu m}.$ The standard deviation is calculated from the  analysis of 10 microspheres. 
In contrast to the first group, the batch of large microspheres displays a much wider size dispersion, as illustrated by Fig.~\ref{fig:5}(c).
Thus, for this group we cannot infer the radii of the specific microspheres employed in the interaction measurements directly from the SEM ensemble characterization. 
We follow instead the correlative microscopy procedure \cite{fonta15,ando18} outlined in the next subsection.

\par Although SEM is not the optimal technique for surface roughness characterization, it is worthwhile to estimate the
two length scales controlling the magnitude of the roughness correction to the Casimir interaction~\cite{vanbree74,maradudin80,czarnecki80,mazur81,bevan1999,zwol08,namsoon17}: the transverse correlation length $\ell_{{\rm C}}$
and the rms roughness amplitude $a$.
The effect of roughness of silica particles is extremely important when probing shorter distances, $L \lesssim 10\,\si{nm}$ \cite{wodka14,valmacco16}.
From images like the one shown in Fig.~\ref{fig:5}(b), we  estimate $\ell_{{\rm C}}<50~\si{nm}$
and $a \lesssim 10~\si{nm}$ for the first set of microspheres, with the highest spikes in the range close to $\sim 100~\si{nm}.$
Since the roughness correction to the Casimir interaction typically scales as $(a/L)^{2}$ \cite{bordag95,klimchitskaya96,Klimchitskaya99},
 we expect corrections $\sim 1\%$ or smaller
 over the distance range probed in our experiment, also because the condition $\ell_{{\rm C}}<L$ reduces the corrections even further \cite{genet03,MaiaNeto05EPL,maianeto05,MaiaNeto06}. 
 Although the presence of long spikes might produce additional roughness corrections beyond the perturbation regime~\cite{broer11,svetovoy15}, 
 they have been neglected when modeling the Casimir interaction
given that our precision is at the level of $\sim 10\%.$ In addition, the good agreement with our data indicates that roughness does not play a major role in our experiment. 
  
Overall, the surfaces of the microspheres from the second set look more smooth in the SEM images. However, they present defects in the form of protrusions and
depressions, as shown in Fig.~\ref{fig:5}(c). 
As they are big enough to be visible 
in the optical microscope, they can be avoided during the experimental runs. 

\subsection{\label{sec:corrmic} Correlative Microscopy}
\par
\begin{figure}
\includegraphics[scale=0.15]{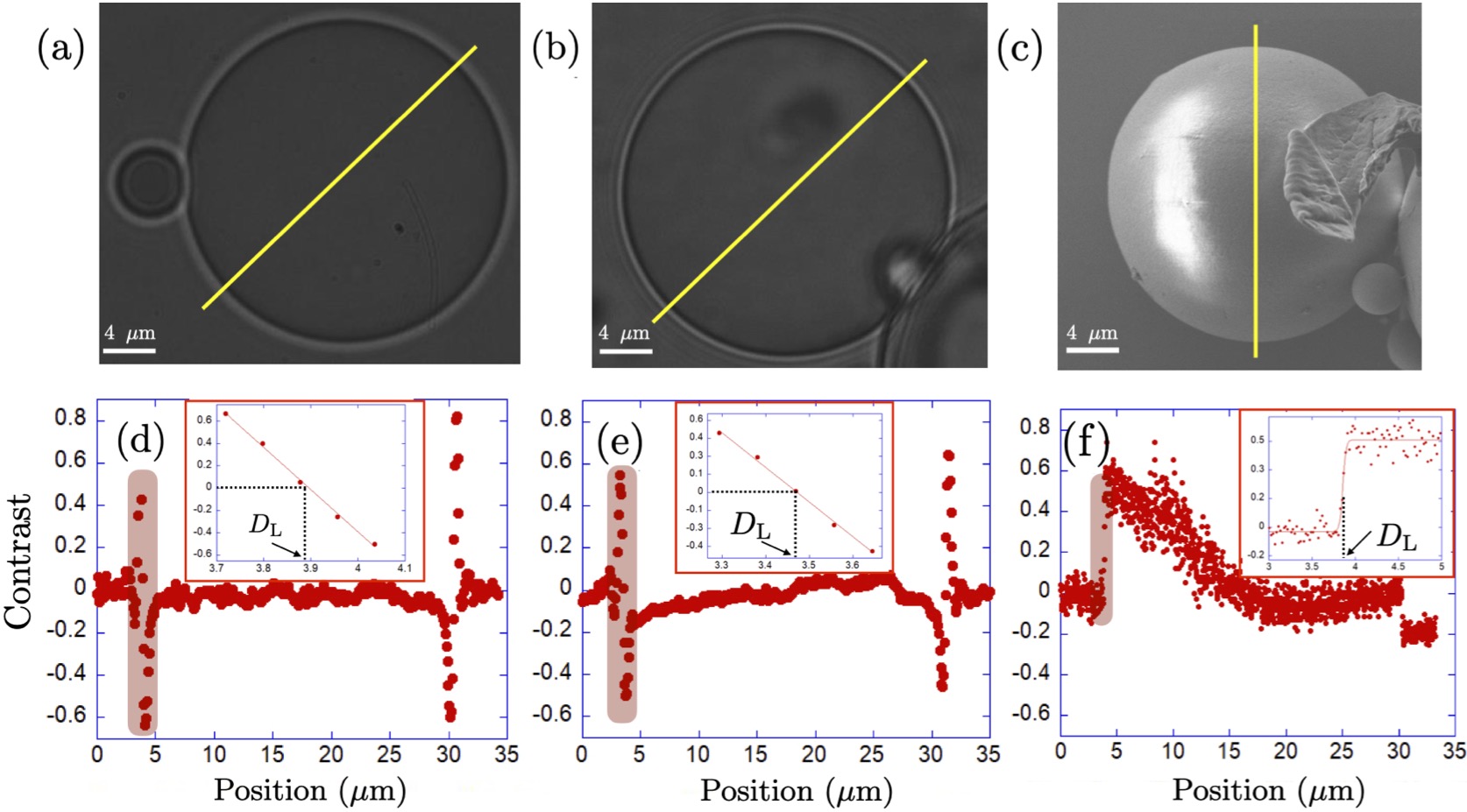}
\centering
\caption{Optical images of the (a) microspheres employed in the interaction measurement and (b) taken as a reference for the method of correlative microscopy. 
The SEM image of the same reference microsphere (b) is shown in (c). Each image is analyzed in terms of the grey level contrast along the diameters indicated by yellow lines. 
(d,e,f)  Contrast versus position for each corresponding image shown on the top line.
The  edge regions (light red) are zoomed in 
the insets.  The edge positions $D_L$ are then determined by appropriate fits (see text for details). }
\label{fig:6}
\end{figure}

 Since it is not possible to carry out SEM on the specific large microsphere used in the experiment, a reliable measurement of its radius is achieved by correlative microscopy \cite{fonta15,ando18}. In a nutshell, this technique consists of inferring the radius of that specific microsphere by correlating optical and SEM images of microspheres from a different sample.
 Figure~\ref{fig:6}(a) shows the optical image of the two microspheres employed in the measurement with low salt concentration.
 Similar images are obtained for the pair of microspheres used in the experiment with added salt. Since the size dispersion of the set of small 
 microspheres is negligible, we focus on the image of the large one. We prepare a sample from the set of large microspheres and  tag one specific microsphere whose 
 optical image, shown in Fig.~\ref{fig:6}(b), is similar to the image of the probed one. In addition, 
we displace the microscope stage along the $z$ axis until its optical image resembles more closely
the image \ref{fig:6}(a)  corresponding to the measurement run. 
 The same microsphere shown in Fig.~\ref{fig:6}(b) is then analyzed with SEM and the resulting image is shown in Fig.~\ref{fig:6}(c). 
 In Figs.~\ref{fig:6}(d,e,f), we plot
 the contrast $C=I/I_{0}-1$ measured along the straight (yellow) lines along the microsphere diameter shown in panels (a,b,c).
They are calculated for each corresponding image shown on top. 
Here, $I$ and $I_{0}$ are the grey levels
at a given image point and at the  background, respectively.

Figure ~\ref{fig:6}(d) shows that the contrast vanishes across the central region and then displays a sharp variation at the microsphere edge, 
which is blurred by diffraction. We define the optical edge as the point where the contrast vanishes inside the edge region highlighted in light red.
The inset is a zoom illustrating the determination of the optical edge coordinate $D_L$ on the lefthand side by a linear fit. 
Such point is not an indication of the actual physical edge~\cite{Gomez2021}. Nevertheless, it is a useful reference that can be measured with an excellent precision from the variation of the contrast.  The same procedure is implemented in Fig.~\ref{fig:6}(e)  for the optical image of the tagged microsphere to be compared with the SEM image.
In both cases, the optical radius is defined as $R_{{\rm opt}}=\frac{1}{2}\left(D_{{\rm R}}-D_{{\rm L}}\right)$, where $D_{{\rm R}}$
is the coordinate of the optical edge on the opposite side of the image.

\par The key requirement in correlative microscopy is to produce a SEM image of the same microsphere that was optically imaged in Fig.~\ref{fig:6}(b).
A rectangular mark made in the central region of the coverslip with a diamond-tipped pen is used as a reference to identify each single microsphere, and in particular
the one selected for the optical image Fig.~\ref{fig:6}(b).  After capturing the optical image, the preparation for SEM is implemented as described in appendix \ref{sec:sem}.

\par The next step is to analyze the SEM image of the tagged microsphere. 
As shown in Fig.~\ref{fig:6}(c), the edge of each microsphere is sharp and not blurred by diffraction. Thus, it is possible to directly access the physical radius $R_2$
from the SEM image. We fit a hyperbolic tangent function to the contrast function plotted in Fig.~\ref{fig:6}(f). 
The inset is a zoom of the edge region revealing the edge position $D_L$ as derived from SEM. 
The physical radius $R_2$ is then determined from $D_L$
and the position of the opposite edge located across the  microsphere diameter. 

The final step consists of correlating the optical and SEM images of the same microsphere in order to derive
the scale factor $r_{c}=R_{{\rm opt}}/R_2.$
The entire procedure is repeated for three distinct tagged microspheres.
After averaging the results, we find $r_{c} = 1.041 \pm 0.005,$
where the uncertainty corresponds to the standard error of the mean. 
Knowing $r_{c}$, the physical radius $R_2$ of a particular microsphere in interaction can be directly inferred by measuring the corresponding optical radius $R_{{\rm opt}}.$
The results for the microspheres employed in the measurement runs are indicated in Sec.~\ref{sec:III}.

\subsection{\label{comparison}Distance upon Contact}

When performing the experiment with high salt concentration, in most experimental runs 
we observed that the probed particle eventually jumps into contact with the 
adhered microsphere. 
By measuring the center-to-center distance $(\bar{X}_2-\bar{X}_1)_{\rm cont}$ upon contact, one can determine the distance upon contact $L_0=(\bar{X}_2-\bar{X}_1)_{\rm cont}-(R_1+R_2)$
thus providing information on the scale of the highest peaks of the rough silica surfaces~\cite{zwol09}. 

Averaging over three contact events, we find  $L_0=(0.2\pm 0.1)\,\mu{\rm m}.$
Such value is compatible with the estimation for the highest asperities based on the SEM images discussed in Appendix \ref{sec:sem}.
The total experimental error is determined by the error of $R_1+R_2$ only, 
as the uncertainty of the center-to-center distance upon contact is much smaller. 

The high peaks associated to the distance upon contact are usually sparse and thus provide a small correction to the Casimir interaction
in the distance range probed in our experiments~\cite{broer11}. Given our limited experimental precision,
we have neglected the contribution of such rare peaks.

\section{\label{sec:detection} Position Detection}

We measure the position of the microsphere center by employing the 
 edge detection algorithm of Refs.~\cite{ueberschar2012novel, yucel2017new}. 
 This method is particularly suitable when considering two microspheres at close distance, since it allows to exclude the region  
 where the images of the microspheres overlap, producing a non-trivial diffraction pattern.
Hence we determine the microsphere edge only within 
the region indicated by the green contour shown in Fig.~\ref{fig:1}(c).
The yellow dots indicate the positions used to fit a circumference, whose center is identified as the position of the trapped microsphere.
 An analogous procedure is implemented for the microsphere adhered to the coverslip.

\begin{figure}[H]
\includegraphics[scale=0.25]{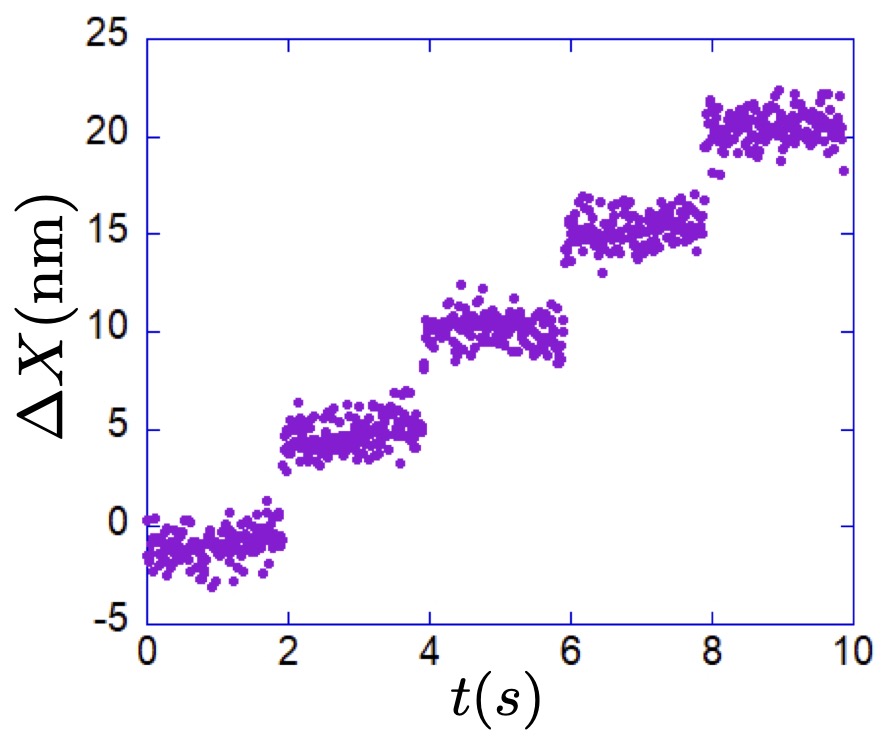}
\centering
\caption{Microsphere position versus time. A silica microsphere is adhered to the cover slip, 
which is driven laterally by $5~\si{nm}$ every 
 $2~\si{s}$ with the help of a piezoelectric nanopositioning system.}
\label{fig:plateaux}
\end{figure}

To test the precision of the position detection method, a silica microsphere of nominal radius $\bar{R}=10\,\mu{\rm m}$ is attached to the coverslip and then 
displaced by nominal steps of $5\, {\rm nm}$ every 
 $2~\si{s}$ by employing a piezoelectric nanopositioning system to drive the microscope stage along the $x$ direction (see Appendix A for details).
 In Fig.~\ref{fig:plateaux}, we plot the 
variation of the microsphere position with time.
The averaged standard deviation for the steps shown in the figure is  $(0.80 \pm 0.04)~\si{nm}$, proving
our ability to detect nanometric displacements on the $xy$ plane.
The average separation
between consecutive steps is $(5.4\pm 0.2)~\si{nm}.$ 
We also validate our position detection method by comparing results for the optical trap stiffness from three different methods: 
position fluctuations, Stokes calibration and modelling based on the Mie-Debye theory of optical tweezers~\cite{Mazolli2003} as discussed in Appendix~\ref{sec:cal}.

\section{\label{sec:env} Environmental Noise}

\par Both the stiffness calibration and the 
interaction measurements critically depend on the assumption of thermal equilibrium \cite{wong06,schaffer07,nicholas14}. Environmental noise such as air-currents, temperature gradients, mechanical vibrations, drifts on the laser and on the microscope stage can easily drive the system far out of thermal equilibrium, creating awkward systematic measurement artefacts. For instance, extra fluctuations on the trapped microsphere are responsible for trap stiffness underestimation; directional drifts can change the relative position of both optically trapped and attached microspheres, drastically changing the interaction among them. As a result, carrying out careful experimental preparation, environmental noise characterization and mitigation actions are crucial to guarantee a thermally limited system 
leading to accurate stiffness calibration and interaction measurements.
\par Let us address each of the environmental noise sources and the actions which have been taken to mitigate them. 
Air-currents and temperature gradients are reduced by symmetrically positioning the air conditioners in the laboratory room, and more importantly covering all the laser optical path, all the optical devices and the optical microscope with a home-made enclosure made of cardboard with a few layers of bubblewrap foil. Mechanical noise is reduced by mounting all of the optical elements in small optical posts attached on an optical breadboard placed on a conventional optical table as discussed in Appendix \ref{sec:system}.
\subsection{\label{sec:drift}Laser Drift}
\par
\begin{figure}
\includegraphics[scale=0.55]{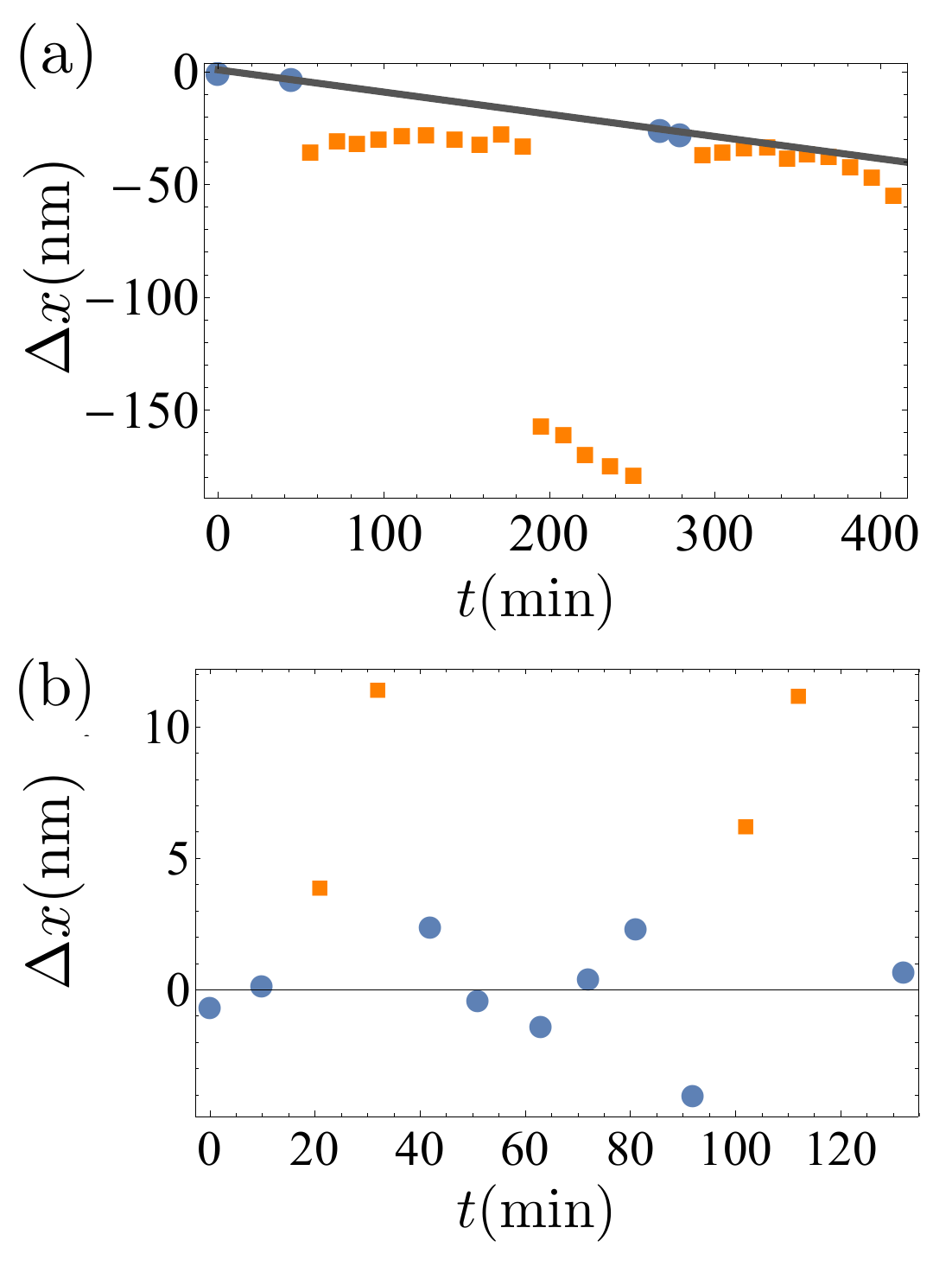}
\centering
\caption{(a) Center of mass position deviation $\Delta x$ of the optically trapped microsphere as a function of time for the experimental runs with (a) low 
and (b) high
salt concentrations. 
Blue data points correspond to 
the situation with no interaction (large distance $D$), while orange ones represent the position deviation in the presence of colloidal interactions. 
The solid line in panel (a) is a linear fit of the blue points indicating a laser drift of  $\sim0.1~\si{nm/min}.$}
\label{fig:drift}
\end{figure}

  Laser drifting is cautiously characterized during all interaction experiments. Figure~\ref{fig:drift}(a) shows
 the long time dependence (temporal scale of $\sim{\rm hours}$) of the position deviation $\Delta x$  from the average initial position of the optically trapped microsphere, when no salt was added to the solution (see section \ref{sec:low}). Each dot represents the mean position of the optically trapped microsphere over an experimental run of $\mathcal{T}=500~\si{s}$, with the exposure time $W=2~\si{ms}$. 
The blue dots correspond to the
 situation with no interaction
as the average separation between the microspheres is  large: $L\gtrsim 500~\si{nm}$ (see Fig.~\ref{fig:1} of the main text). 
On the other hand, the orange plateaux
correspond to distances $L<400~\si{nm}$ such that the interaction is non-negligible. 
The figure shows that as the microspheres are moved closer and further away, the optically trapped microsphere has not returned back to its original position, being slowly carried by a laser drift. Repeating this back and forth protocol and fitting a straight line to the set of isolation points, we find a laser drift of $\sim0.1~\si{nm/min}$, which was used as a correction for the unperturbed equilibrium position $L_{\rm eq}^{\rm opt}$ (see section \ref{sec:low}) in the experiment with low salt concentration. 

\par In Fig.~\ref{fig:drift}(b), $\Delta x$ as a function of time is shown for the 
experiment with high salt concentration (see section \ref{sec:high}). 
In the case of interaction (orange points),
the position deviations are along the positive $x$ axis, indicating attraction as expected (see Fig.~\ref{fig:1}(a)).   
In contrast to the low-concentration experiment shown in Fig.~\ref{fig:drift}(a), here no substantial laser drift takes place, as indicated by the 
blue dots corresponding to large distances $L\gtrsim 500~\si{nm}.$
We hypothesize that this behavior is due to different environmental conditions between the two experimental rounds, since they were conducted on distinct days. 
As indicated by Fig.~\ref{fig:drift}(b), the blue dots are scattered around zero with a standard deviation
 $\sigma_{{\rm drift}}\sim2~\si{nm}$, which allows us to estimate the experimental error in our force interaction measurements.
  For a trap stiffness $k\sim1~\si{fN/nm}$, we then have a force error $\delta F\approx k\sigma_{{\rm drift}}\sim2~\si{fN}$, which is shown in Fig.~\ref{fig:4} as a shaded region and also represents the minimum detectable interaction force of our experimental setup.
  
\subsection{\label{sec:stage_drift}Microscope Stage Drift}

\par Microscope stage drifts were mitigated by active control stabilization using a feedback loop \cite{mcgorty13} simultaneously in the three $x$, $y$, and $z$ axis. As shown in Fig.~\ref{fig:1}(b) and described in section \ref{sec:system}, the attached microsphere's position is continuously monitored by a CCD camera taking position measurements at a sampling rate of $32~\si{fps}$ and averaging over $20$ frames, thus giving an effective sampling rate of $r_{{\rm feed}}=1.6~\si{fps}$. Each measurement is then returned back as a feedback signal to the digital piezoelectric controller, which performs the position corrections, thus enabling to keep the position variations along the three axes below $10~\si{nm}$ for each experimental run of $500~\si{s}$.
\subsection{\label{sec:cal}Trap Stiffness Calibration}
\par Stiffness calibration is carried out by applying the two well-known methods of potential energy fitting \cite{florin1998,nicholas14,Gieseler2021} and drag force counterbalance \cite{neuman04,nicholas14}. In the former protocol, which is entirely based on thermal fluctuations analysis \cite{norrelykke10,norrelykke11,Melo20}, the energy potential landscape of an optically trapped microsphere is constructed by monitoring the diffusive Brownian motion of the microsphere along the $y$ direction, which is orthogonal to the 
interaction force as indicated by Fig.~\ref{fig:1}(c). The corresponding trap stiffness is then obtained by fitting a quadratic function to the potential energy. In the latter protocol, the optically trapped microsphere experiences a viscous drag force by creating a fluid flowing in its surroundings. By counterbalancing that hydrodynamical force, which is described by the Stokes-Fax\'en law \cite{berg-sorensen04,tolic06,schaffer07,happel83}, to the optical tweezers' restoring force, the desired stiffness along the direction of the fluid flow is obtained.
\par
\begin{figure}[t]
\vspace*{0.5cm}
\includegraphics[scale=0.21]{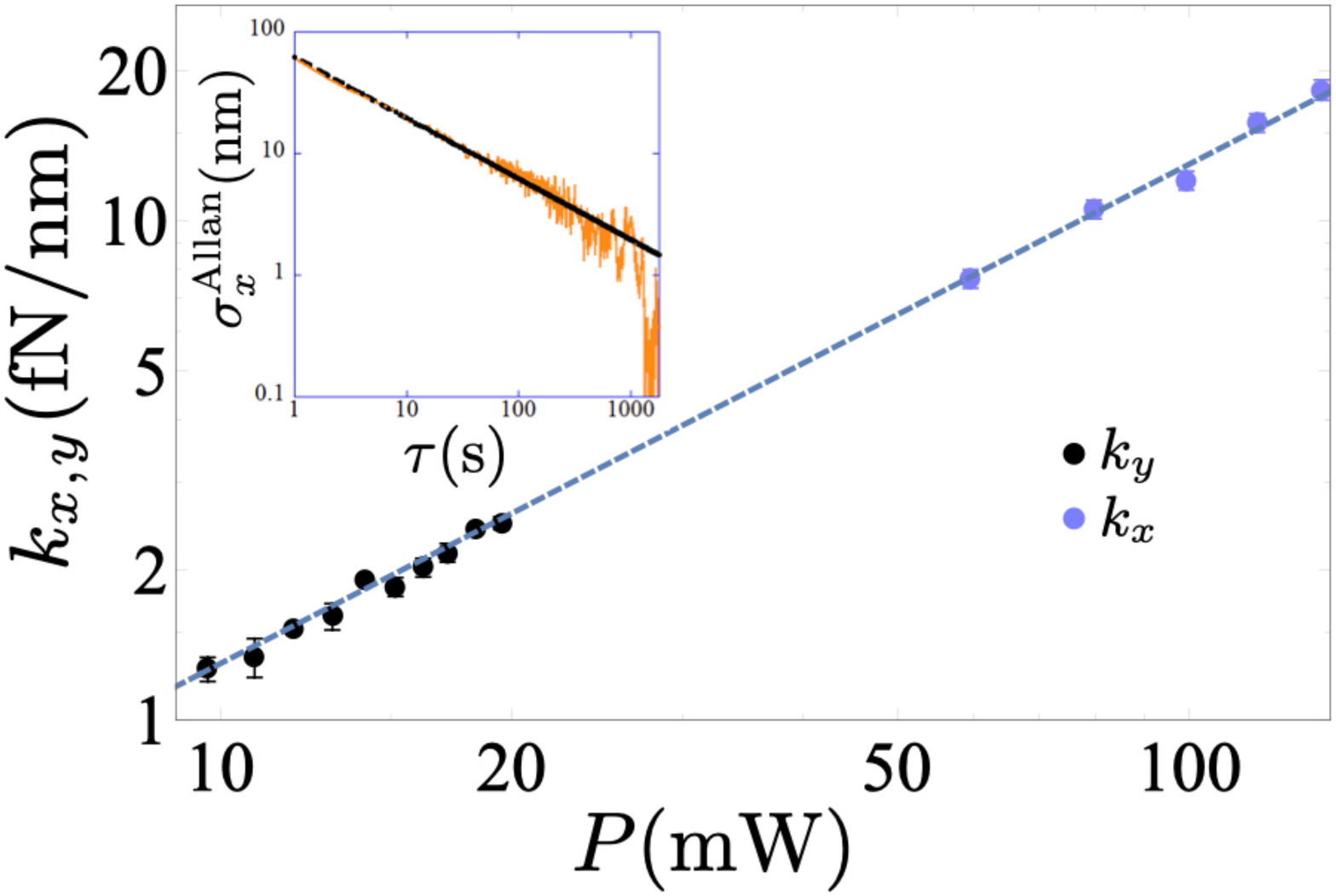}
\centering
\caption{Optical tweezers' stiffnesses $k_{x}$ (blue dots)  and  $k_{y}$ (black dots)  versus laser power $P$ at the objective entrance port.
 $k_x$ and $k_y$ are measured by the drag force and fluctuation methods, respectively.  
 While $k_x$ is measured for an isolated microsphere, $k_y$ is obtained from the fluctuations in actual interaction measurement runs. 
 We find the same angular coefficient when fitting the two sets of data points separately, which also agrees with an independent theoretical calculation. 
 The resulting best fit (dashed line) indicates the equivalence between the two calibration protocols, as well as the isotropy of the optical force field on the $xy$ plane as expected for circular polarization. Inset: Allan deviation of an isolated optically trapped microsphere's $x$ position as function of time. The dashed line represents the thermal limit deviation showing that the system is thermally limited along the whole time interval until $\tau\sim1000~\si{s}$.}
\label{fig:stiffness}
\end{figure}

 In Fig.~\ref{fig:stiffness}, we plot the trap stiffness as a function of the laser power $P$ at the objective entrance port. 
We show the results from the two calibration protocols in the same plot. Each protocol is employed for an individual silica microsphere 
(nominal radius $\bar{R}_{1}=2.5\,\mu{\rm m}$) from the same batch (see Appendix \ref{sec:sample} for details). The size dispersion within a given batch is negligible, so we expect the results for the two measured microspheres
 to agree within error bars.
The black dots are the trap stiffness $k_{y}$ along the $y$ axis obtained from the analysis of the Brownian motion of the optically trapped microsphere in actual 
interaction measurement runs. This motion is monitored with an exposure time of $W=2~\si{ms}$ and with a sampling rate of $f_s=10~\si{fps}$. Since the correlation time of 
Brownian fluctuations in the optical
trap is $\tau_{{\rm C}}\sim100~\si{ms}$,
those values for $W$ and $f_s$  respectively avoid motion blur corrections \cite{wong06,vanderHorst10} and correlated data \cite{flyvbjerg89}. Fitting a linear function crossing the origin to the set of calibration points gives the angular coefficient $k_{y}/P=(0.129\, \pm\, 0.001)~\si{fN\,nm^{-1}\, mW^{-1}}$, where the uncertainty is obtained from the weighted fitting procedure. 

\par The blue dots in  Fig.~\ref{fig:stiffness} are the results for the trap stiffness $k_{x}$ along the $x$ axis obtained by the drag force calibration method. 
They are acquired with $W=2.8~\si{ms}$ and $f_s=357~\si{fps}$, which guarantee proper relaxation and counterbalance monitoring. The corresponding weighted linear fit to this calibration set yields the angular coefficient $k_{x}/P=(0.129\, \pm\, 0.003)~\si{fN\,nm^{-1}\, mW^{-1}}$, which is in excellent 
agreement with the value obtained for  $k_{y}/P$ from Brownian fluctuations. 
For convenience, we have employed 
 circularly-polarized trapping beams in all experiments reported in this paper. Thus,
we expect the optical trap stiffness to be the same along all directions on the $xy$ plane, also because the effect of residual astigmatism on the optical force field is negligible in this size range~\cite{dutra12,dutra14}. 
Indeed, isotropy of the optical force field on the $xy$ plane allows us to use the Brownian fluctuations along the $y$-direction, which is orthogonal to the 
interaction force between the two microspheres,  as a check of the optical force during the interaction runs.
The dashed straight line shown in Fig.~\ref{fig:stiffness}, corresponding to the best fit linear function, 
provides further visual indication of the 
compatibility between the two calibrations protocols performed along the orthogonal $x$ and $y$ axis.

\par As a third method, 
we have employed the Mie-Debye theory of optical tweezers~\cite{Mazolli2003}
to calculate the optical trap stiffness.
Such absolute calibration method~\cite{dutra12,dutra14} also relies on an independent characterization of all input parameters required by the theoretical model: the objective transmittance, beam waist at the objective entrance port, 
microsphere radius, and refractive indexes of the microsphere and host medium.
 In addition, the known values for the laser wavelength, objective numerical aperture 
 and focal length ($f=4.44\,{\rm mm}$) are also required (see Appendix~\ref{sec:system}). 
 
\par To determine the laser power at the sample region, 
the objective transmittance is characterized as in Ref.~\cite{viana2006characterization}:
a laser beam, with the same waist employed in the trapping experiment, 
is transmitted through the water-immersion 
objective and then reflected back into the objective by a mirror attached to the microscope stage. 
We found a single-pass transmittance of $0.55\pm 0.06$. The method of Ref.~\cite{viana2006characterization} also allows for the 
characterization of the beam waist $w$ at the objective entrance port. We found $w=2.34\,{\rm mm}.$

\par We measured the radius of the trapped silica microspheres from SEM images and found $R_{1}=(2.35\pm 0.02)~\si{\mu m}$
with negligible size dispersion within our batch
as discussed in Appendix~\ref{sec:sem}.
Two independent methods were recently employed to measure 
 the refractive index of silica microspheres from the same batch 
 at $\lambda_0=470\,{\rm nm}$~\cite{Gomez2021}.
The result is smaller than 
the bulk fused silica index, which we attribute to the porosity of the beads (see also \cite{Blakemore2019} for related findings from a mass measurement).
 In order to infer the refractive index at the trapping laser wavelength $\lambda_0=1064\,{\rm nm},$
we employ the (Mie-based) extended Maxwell-Garnett (EMG) effective medium theory \cite{Doyle1989, Rupin2000}
assuming that our silica beads are filled with empty pores. From the refractive index measured at $470\,{\rm nm},$
we obtain a volume filling fraction $0.078,$ which we then employ to calculate $n_{\rm bead}=1.4146\pm 0.0019$ from 
the bulk fused silica refractive index $n_{\rm silica}=1.4496$ at $\lambda_0=1064\,{\rm nm}$ \cite{Malitson1965} by employing again the EMG theory. 
At this wavelength of interest, 
the refractive index of distilled water at $24\,{}^{\circ}{\rm C}$ is $n_{\rm water}=1.3242$~\cite{Daimon2007}.
We take this value as the refractive index of the host medium, since its modification for a salt concentration of $\sim 0.2\,{\rm mM}$ (see Sec.~IIIA)
is negligible ($\Delta n\sim 10^{-5}$ according to \cite{Tan2015}).

\par  We assume that optical effects of refraction at the planar interface
between the glass slide and the sample region 
are canceled when using the water-immersion objective (see Appendix~\ref{sec:system}).
Most importantly,  the spherical aberration introduced by the interface, which might
lead to a strong reduction of the trap stiffness~\cite{Viana2007}, is corrected by the water-immersion objective.
We disregard modifications due to a possible small astigmatism of the trapping beam, which are typically 
unimportant for radii $R > \lambda_0$~\cite{dutra12,dutra14}. Finally, we also neglect 
the optical reverberation between the microsphere and the glass slide~\cite{dutra2016theory} given the large distance ($\approx 10\,\mu{\rm m}$) between the 
trapped particle and the slide. 
We then find $k_{x}/P=k_{y}/P=(0.124\, \pm\, 0.018)~\si{fN\,nm^{-1}\, mW^{-1}}$, in very good agreement with the measurements discussed above. 
Here, approximately $ 80\%$ of the theoretical error
originates from the objective transmittance measurement, with
the refractive index of silica and radius uncertainties accounting for the rest.

\par Overall, the agreement between the three different calibration methods provides further validation of our detection method and
 indicates that non-thermal
fluctuations in our system are negligible given the precision of our measurements.
The latter is an essential requirement for probing the interaction potential from 
the microsphere position fluctuations. We provide additional evidence that our system is thermally limited 
by analyzing the Allan deviation in the next sub-section.

 \medskip
\subsection{\label{sec:allan}Allan Deviation Stability Analysis}

\par A successful Brownian stiffness calibration 
already indicates that the motion of the optically trapped microsphere is thermally limited. However, it is also important to determine the optimal time over which it remains in that state, \emph{i.e.} in the absence of any extra non-thermal fluctuations and drifts. One way to quantify this stabilization time is performing an Allan-deviation stability analysis \cite{allan66,czerwinski09,landsdorp12,hebestreit18b,schnoering19}. The Allan deviation of the position of the optically trapped microsphere along the $x$ axis is defined as 
\begin{eqnarray}
\sigma^{{\rm Allan}}_{x}\left(\tau\right)=\left\{\frac{1}{A-1}\sum_{j=1}^{A-1}\frac{1}{2}\left[\langle x_{j+1}\rangle\left(\tau\right)-\langle x_{j}\rangle\left(\tau\right)\right]^{2}\right\}^{1/2}\,,\nonumber\\
&&
\end{eqnarray}
where $A=\mathcal{T}/\tau$ is the number of independent length-$\tau$ blocks in a run of total duration $\mathcal{T}.$
The  $j^{{\rm th}}$ bin-average $\langle x_{j}\rangle\left(\tau\right)$ is 
taken over the interval $\left[(j-1)\tau,j\tau\right]$ with $f_s\tau$ points,
where  $f_s$ is the sampling rate.
The Allan deviation measures the average position variation among consecutive temporal intervals of length $\tau$. In the absence of non-thermal noise, the longer the interval $\tau$, the closer the averages in consecutive intervals will be, thus gradually decreasing the Allan deviation with increasing $\tau$. 
\par In the inset of Fig.~\ref{fig:stiffness}, 
we plot the Allan deviation  $\sigma^{{\rm Allan}}_{x}\left(\tau\right)$ versus time $\tau$ for an optically trapped microsphere under the 
 environmental conditions corresponding to
 the Casimir measurement with high salt concentration
 (see Fig.~\ref{fig:drift}(b)). The surface-to-surface distance is 
 $L\sim 800\, {\rm nm},$ the total time duration is 
  $\mathcal{T}=3200~\si{s}$ and the sampling rate is  $f_s=1~\si{Hz}.$
The dashed line shows the thermal limit deviation $\sigma^{{\rm Allan}}_{{\rm Th}}\left(\tau\right)=\sqrt{2\gamma k_{B}T/k_x^{2}\tau}$ \cite{czerwinski09}, 
where all parameters are determined independently of the Allan deviation analysis, with no fitting procedure. 
We take $k_{x}=1~\si{fN/nm}$, obtained from the calibration procedure described in the previous sub-section, the Stokes friction coefficient $\gamma=5\times10^{-7}~\si{kg/s}$ corrected for the proximity to the coverslip \cite{happel83}, and the measured temperature ${T}=(296\pm 1)~\si{K}$. 
The blind comparison between the experimental results and $\sigma^{{\rm Allan}}_{{\rm Th}}\left(\tau\right)$
shows that our system is stable and thermally-limited out to $\approx 1000~\si{s}$,
 approximately twice the time interval of all experimental calibration and interaction runs. 
For times $\tau\gtrsim  1000~\si{s}$, lack of statistics hinders any analysis. 


\begin{thebibliography}{141}
\bibitem [{\citenamefont {}\ and\ \citenamefont
  {}()}]{foot}%
  \BibitemOpen
   {\bibinfo  {journal} {For the specific pair of runs shown in the inset of Fig.~\ref{fig:2}, it is more convenient for clarity to 
define the relative potential offset by
taking $L_{{\rm ref}} = 0.40~\si{\mu m},$ since the distances are larger than those shown in the main plot}\
  }\BibitemShut {NoStop}%
\end{thebibliography}
\end{document}